\documentclass[review]{elsarticle}

\usepackage{lineno,hyperref}
\usepackage{blindtext}
\usepackage{amsmath,amsfonts}
\usepackage{booktabs}
\usepackage{float}
\usepackage{multirow}
\usepackage{subcaption}
\usepackage[dvipsnames]{xcolor}

\journal{Experimental and Computational Multiphase Flow}

\biboptions{authoryear}

\ifpdf
    \graphicspath{{Figs/Raster/}{Figs/PDF/}{Figs/}}
\else
    \graphicspath{{paper/Figs/Vector/}{paper/Figs/}}
\fi
\usepackage{atbegshi}
\AtBeginDocument{\AtBeginShipoutNext{\AtBeginShipoutDiscard}}

\let\oldeq\equation{}\def\equation{\par\vspace{-\parskip}\oldeq}

\begin{document}
\begin{frontmatter}

\title{A conservative level set method for liquid-gas flows with application in liquid jet atomisation}

\author[mymainaddressA,mymainaddressB,mymainaddressC]{Panagiotis Lyras},
\address[mymainaddressA]{MultiFluidX, Grigoriou Afxentiou 93, 15770 Athens, Greece}
\address[mymainaddressB]{Lyras LP, 24133 Kalamata, Greece}\cortext[mycorrespondingauthor]{Corresponding author} 
\address[mymainaddressC]{School of Electrical and Computer Engineering, National Technical University of Athens, 15773, Zografou, Greece} 
\author[mymainaddressA,mymainaddressB]{Antoine Hubert}
\author[mymainaddressA,mymainaddressB]{Konstantinos G. Lyras\corref{mycorrespondingauthor}}
\ead{k.lyras@multifluidx.com}

\begin{abstract}
In this paper, a methodology for modelling two-phase flows based on a conservative level set method in the framework of finite volume method is presented. The novelty of the interface capturing method used here lies on the advection of level set which is solved with a WENO scheme and is corrected with a novel re-initialisation method for retaining its signed distance function character. The coupling with the volume of fluid method is done with a simple algebraic approach and with the new algorithm the accumulated mass conservation errors remain reasonably low.
The paper presents a unique coupling between the level set method and the Eulerian-Lagrangian-Spray-Atomisation approach for modelling spray dispersion in liquid atomisation systems. 
The method is shown to have good accuracy providing similar results to other numerical codes for the classical tests presented. 
Preliminary results are also shown for three-dimensional simulations of the primary break-up of a turbulent liquid jet obtaining results comparable to direct numerical simulations. Consequently, the coupled method can be used for simulating various two-phase flow applications offering an accurate representation of the interface dynamics.
\end{abstract}


\end{frontmatter}

\section{Introduction}
\subsection{Scope}
Among the most popular methods for capturing the interface between two immiscible fluids are the volume of fluid (VOF) and the level set (LS) methods \citep{Prosperetti2009}. The interface is captured advecting a marker function as initially proposed by \citep{Hirt1981}, which defines the cells that belong to one phase or another. The advection method for the marker function is particularly important and is most of the times a fundamental determinant that differentiates the various proposed methods. 
 
VOF method is a popular implicit method for capturing the interface between two fluids advecting the volume fraction of one fluid in a computational cell, using, among others, an algebraic or geometric method which can be formulated for both incompressible and compressible flows with or without phase change \citep{Hirt1981,Lafaurie1994,Scardovelli1999,Gueyffier1999,Scardovelli2000,Aulisa2003,Tryggvason2011,
Agbaglah2011,Fuster2009,Pozzetti2018}. The method has been proved to be very accurate and has been tested for different multiphase systems. Besides, it is by nature mass conserving. In the LS method, the signed distance function is calculated defining different levels that can be either positive or negative, with the interface located at the zero level. The method is highly accurate for calculating the position of the interface and curvature with multiple applications \citep{Osher1988,Chang1996,Sethian1996,Sethian1999,Enright2002,Osher2006,Tanguy2005,Lakdawala2014,Arienti2014,Balcazar2015,Sandberg2019}
Both methods can adjust naturally to any shape and describe the changes in the fluid shape under deformation, rotation or translation. In the context of an Eulerian grid, the interface in VOF methods is located by the cells with volume fraction values between 0 and 1, whereas in LS methods the transition from one fluid to another occurs in more layers of cells, usually with a predefined narrow band \citep{Osher2006}. 
The standard LS function usually tends to generate unphysical volume changes. The induced spurious oscillations affecting the solution accuracy should be avoided and the interface thickness and velocity profile should be kept constant \citep{Olsson2005}. 
Attempts for addressing the mass conservation of LS methods have been made in various studies for using level set in Navier-Stokes solvers for incompressible two-phase flows with surface tension, leading to derivations that include to some extent a coupling with VOF \citep{Bourlioux1995,Sussman2000}. These coupled approaches require the advection of both the volume fraction and distance function for updating the displacement of the interface between two fluids, conserving mass while demonstrating reasonable accuracy. 
Many different methods that couple volume of fluid with let set have been proposed since \citep{Wang2009,Balcazar2014,Zhao2017}. These usually employ a geometrical reconstruction of the interface such as the piecewise linear interface calculation (PLIC) scheme. Other approaches than PLIC for coupling VOF with LS have been also proposed in \citep{Kees2011,Ferrari2017,Haghshenas2017}. 
\cite{Olsson2007} proposed a re-initialisation equation in order to restore the level set function back to its hyperbolic tangent shape, which is solved in steady state. \cite{Balcazar2014} developed a formulation based on the standard level set that conserves mass, using a PLIC method for reconstructing the interface. The re-initialisation of the level set function was based on the method proposed by \cite{Olsson2007} using a compressive term for the cells close to the interface. The method was developed in the framework of finite volume method and was shown that it can be used for two-phase flows when using unstructured meshes. \cite{Albadawi2013} proposed a finite volume approach that solves a transport equation of the volume fraction that is then mapped into the signed distance function that is not advected directly solving a level set equation. Improvement in curvature calculation was reported with overall good accuracy for the presented tests. 
\cite{Dianat2017} developed a coupling of the level set and volume of fluid methods employing the concept of the area of fluid for advecting the liquid volume fraction using an iterative clipping and capping algorithm. Both the LS and VOF functions are advected by solving a transport equation for each one of them: the volume of fluid was advected employing an interface compression scheme whereas the LS function used a van Leer TVD scheme. 
A different LS-VOF coupling suitable for overlapping and moving structured grids was proposed in \cite{Zhao2017} using a geometric VOF method for advecting the volume fraction. In this PLIC scheme, the interface was advected using a hybrid split (Eulerian implicit-Lagrangian explicit interface advection) which was proven to be very accurate for the tests performed.   
Various codes, specifically for spray and liquid atomisation modelling have been developed based on novel ideas for coupling VOF and LS methods. \cite{Menard2007} used the level set method for tracking the interface, together with the ghost fluid method for capturing sharp discontinuities. The authors used a projection method for solving Navier-Stokes equations, coupled with a transport equation for the level set function following the method of \cite{Sussman2000} for ensuring mass conservation. The method showed excellent capabilities for simulating primary atomisation and describing the complex phenomena that cause the jet disintegration. In order to study the liquid jet atomisation, \cite{Chesnel2011} used a similar approach to \cite{Menard2007} to perform direct numerical simulations using a coupled LS and VOF formulation. The so-called Eulerian-Lagrangian-Spray-Atomisation (ELSA) method of \cite{Vallet2001} was implemented for studying the atomised jet close to the injector. Similarly, \cite{Duret2013} have used a similar level set method with ELSA for solving the surface density ($\Sigma$) equation for modelling liquid atomisation. The code employed a fifth-order WENO scheme for solving the derivatives and was used to estimate the contribution of the turbulence source term in the $\Sigma$-equation. The latter, offers a strong modelling tool for simulating sprays and atomisation process and can be easily coupled with interface capturing methods regardless the interface reconstruction method used. This idea was used by \cite{NavarroMartinez2014} using a stochastic method for calculating the sub-grid fluctuations of the surface density and the liquid volume fraction, implementing ELSA within the second-order code Boffin.
Similar to the methods in \citep{Sussman2000,Menard2007,Duret2013} the hybrid approaches employ the interface normals calculated from the level set, in the interface reconstruction used by the VOF function using a 6 or 9 points stencil for the interface reconstruction.

A novel coupling of the Eulerian-Lagrangian-Spray-Atomisation model with the level set method is implemented in this study for modelling multiphase flow systems that involve droplets, bubbles, and other flow regimes that arise in free-surface flows such as liquid-gas systems. This is done by using a level set method which employs a specific numerical scheme for re-initialising the level set function which has been recently shown to give a better accuracy than the schemes usually deployed for this purpose \citep{Lyras2020}. 
The level set advection differs from other similar works such as in \cite{Dianat2017} since we use a WENO scheme for solving the equation for level set. 
For simplicity, the mapping from VOF to level set is done assuming that the interface is located at the 0.5 contour of the volume fraction. The calculated level set function is then used for calculating the surface tension force into the momentum equation and for updating the two-phase mixture properties.
Our method here also fundamentally differs from other related works as in \cite{Albadawi2013,Lyras2020} since contrary to those coupled LS-VOF methods, the LS field is advected here by solving the level set equation at the interface. An interesting feature of the method here, is that the interface is described without the need of either a high-order scheme for solving for the liquid volume fraction or a geometrical method as in \cite{Menard2007,Lyras2020} that would demand an interface reconstruction method and that would increase the computational cost and complexity of the numerical algorithm.
This novel algorithm for solving for level set, provides a simpler formulation compared to PLIC-based methods as the one in \cite{Menard2007}, offering an accurate representation of the interface for both structured and unstructured meshes. 
   Significantly, the ELSA model which is mostly used as a standalone model for modelling liquid atomisation in the literature or is coupled with VOF methods, deploys here the level set method for improving the interface capturing and consequently the spray dispersion predictions. Although some scarce numerical works with ELSA and the LS method are present in the literature, these are based on geometrical methods for advecting the VOF field such as in \cite{Duret2013} with the associated drawbacks mentioned above. Additionally, neither uses the re-sharpening algorithm for the re-distancing of the LS field that we employ here and which has been proven to improve the accuracy of the calculated volume fraction for long-time simulations and simulations with coarse grids, avoiding accumulated mass conservation errors \citep{Lyras2020}.
  
The presented methodology is first tested against published results for both structured and unstructured meshes. Comparisons of the coupled LS-VOF through different numerical tests revealed that the method improves the accuracy of solution while conserving mass. Results for primary atomisation are also presented showing that the method can model liquid jet atomisation and can be used for calculating liquid volume fraction and surface density.  
The methodology is implemented in the code MPflow, which is a numerical solver based on OpenFOAM \citep{Weller1998} and is a dedicated multiphase flow with various capabilities such as turbulent mixing, sprays, dispersion and combustion. The code is second-order in space and is constructed for computational cells of arbitrary shape. 
    
\section{Methodology}
\subsection{Signed distance function and volume fraction advection}
The interface which separates two fluids is represented here by the level set function $\psi(\mathbf{x},t)$. Depending on whether a given point $(\mathbf{x},t)$ with a distance $d$ to the interface $\Gamma$, $d=min|\mathbf{x}-\mathbf{x}_{\Gamma}|$, exists in one fluid or the other, $\psi(\mathbf{x},t)$ is defined as $\psi(\mathbf{x},t)=+d$, or $\psi(\mathbf{x},t)=-d$, respectively. The interface is then defined as the set of points that belong to the zero-level
\begin{equation}
\Gamma=\left\{\mathbf{x}| \psi(\mathbf{x},t)=0 \right\}
\end{equation}
The level set function is then a distance function that is defined wherever an interface exists. The distance function can be advected using the following
\begin{equation}
\frac{\partial \psi}{\partial t}+ \mathbf{u} \cdot \nabla{\psi} =0
\label{psiEqn}
\end{equation}
where $\mathbf{u}$ is the velocity field. The above equation can be solved using any high order scheme for hyperbolic systems of the ENO (essentially non-oscillatory) schemes family or the Runge-Kutta method \citep{Liu1994}. A similar advection equation is used for the marker function in volume of fluid methods.  
The advection of $\psi$ and its re-initialisation in order to retain a distance function, even with high-order schemes or for divergence free velocity fields, lead to mass conservation errors and consequently the volume which is bounded by the zero level is not conserved. To represent density and viscosity discontinuities over the interface, instead of $\psi$, a Heaviside function $H$ is used which is based on 
\begin{equation}
H(\psi) 
= \begin{cases}
1 & \text{\quad if in fluid 1}  \\
\in (0,1) & \text{\quad at the interface} \\
0 & \text{\quad if in fluid 2}  
\end{cases}
\label{HEqn}
\end{equation}
Then the mass conservation requires the following to be conservative for incompressible flows 
\begin{equation}
\frac{\partial H(\psi)}{\partial t}+ \mathbf{u} \cdot \nabla{H(\psi)} =0
\label{HpsiEqn}
\end{equation}
The mass error is no-longer zero when time $t>0$, accumulating during the simulation. Choosing an appropriate value for the Heaviside function at the interface is extremely important for conserving mass since the total mass of the fluid depends on the expression for $H$. The approach presented here, calculates the appropriate function $H$ based on the updated signed-distance function. The staring point for this update is mapping the volume fraction to the initial guess for the level set function assuming the interface is located at the $\alpha-0.5$ isosurface. 

\subsection{One-fluid flow approach}
The ultimate goal of this paper is to propose an Eulerian-Lagrangian approach for simulating liquid atomisation and two-phase spray flows. In the Eulerian framework, the spray is considered as a continuum across the domain, whereas in the Lagrangian description, the liquid fragments such as ligaments, droplets are modelled as a discrete phase considering their mutual interactions. 
The incompressible Navier-Stokes equations without any phase change phenomena following \cite{Prosperetti2009} are given using Einstein's notation 
\begin{equation}
\frac{ \partial u_{j}}{\partial x_{j}}= 0 
\label{massEqn}
\end{equation}

\begin{equation}
\rho \left[ \frac{\partial u_{i}}{\partial t}+ u_{j}\frac{\partial u_{i}}{\partial x_{j}} \right] = -\frac{\partial p}{\partial x_{i}}+ \frac{\partial}{\partial x_{j}}\left( 2\mu S_{ij} \right) - \mathbf{g}\cdot\mathbf{x}\nabla \rho + F_{\sigma}
\label{momentumEqn}
\end{equation}
where $u_{i}(x, t)$ represents the $i$-th component of the fluid velocity at a point in space, $x_{i}$, and time, $t$. Pressure and density are denoted with $p$, $\rho$, the magnitude of the gravitational acceleration is $g=9.81m/s^2$ and the strain rate tensor is  $S_{ij} = \frac{1}{2}(\frac{\partial u_{i}}{\partial x_{j}} + \frac{\partial u_{j}}{\partial x_{i}})$. Since the incompressible case is considered here, $S_{kk}=0$. The last term $F_{\sigma}$ in Eq.\eqref{momentumEqn} is localised on the interface and corresponds to the surface tension force. The volume fraction is calculated first as, 
\begin{equation}
\frac{\partial \alpha}{\partial t}+ \mathbf{u} \cdot \nabla{\alpha} =0
\label{alphaEqn}
\end{equation}  
In its integral form Eq.\eqref{alphaEqn} is written as
\begin{equation}
\int_{\Omega}^{} \frac{\partial \alpha}{\partial t} \text{d}\Omega + \int_{\partial \Omega}^{} \alpha \mathbf{u}\cdot \mathbf{n}\text{d}S = 0 
\end{equation}    
where $S, \mathbf{n}$ are the surface of volume $\Omega$ and the normal vector. The discretised form of this equation can be given by a forward Euler scheme as 
\begin{equation}
\frac{\alpha^{n+1}-\alpha^{n}}{\Delta t}\Delta \Omega + \sum_{f}(F^{n}_L + \lambda F^{n}_{H}) = 0\label{alphaEqndistrite}
\end{equation}
The right hand side of Eq.\eqref{alphaEqndistrite} is the sum over all face $f$ of the volume $\Omega$ at the previous iteration. The advective fluxes $F_L$  and $F_H$ are calculated from the volume flux at the face $f$. The parameter $\lambda$ varies from 1 at the interface to 0 in cells with $\alpha=$ 1 or 0. Boundedness of the temporal solution can be achieved via face value limiters, by limiting the face fluxes or among others by using TVD schemes which enable prediction of sharp changes in field values that preserves monotonicity (no spurious oscillations in the solution) for different types of grids.
More details for the calculation of the fluxes $F_L$  and $F_H$ are given in \cite{Deshpande2012}. 
The level set function is then solved via Eq.\eqref{psiEqn} and the distance function is advected. The level set is re-initialised for allowing $\psi$ to remain a signed distance function, which is done solving the Eikonal equation $|\nabla \psi| = 1$. 
\begin{equation}
\frac{\partial \psi_{d}}{\partial 
\tau}=\textrm{sgn}(\psi)(1-|\nabla \psi_{d}|)
\label{reinitGeneric}
\end{equation}
The new corrected distance function, $\psi_{d}$, is calculated using some iterations using the solution $\psi$ as an initial guess in the re-initialisation step, $\psi_{d}(t=0)=\psi$. The distance function is corrected locally in a narrow band surrounding the interface with thickness 2$\epsilon$. Outside this narrow band the Heaviside function is formulated as in Eq.\eqref{HEqn}. In the vicinity of the interface, $H$ becomes for $ -\epsilon \leq \psi \leq  \epsilon$ \citep{Olsson2005,Tryggvason2011} 
\begin{equation}
H(\psi) = 
\frac{1}{2}\left[ 1+\frac{\psi}{\epsilon}+\frac{1}{\pi}sin(\frac{\pi \psi}{\epsilon})\right] 
\end{equation}
The two fluids are treated as one fluid with properties that change across the interface.
The pseudo-fluid properties can be then calculated as
\begin{equation}
\rho(\psi) = \rho_1 H + \rho_2 (1-H) 
\end{equation}
\begin{equation}
\mu (\psi) = \mu_1 H + \mu_2 (1-H)
\end{equation}
Using the curvature correction from $\psi$, the surface tension force acting on the interface is evaluated similarly to \cite{Brackbill1992}
\begin{equation}
F_{\sigma}=\sigma \kappa \delta(\psi)\nabla \psi 
\end{equation}
where $\delta()$ is the Dirac function. $F_{\sigma}$ is considered for a small narrow band near the interface with the user defined thickness $\epsilon$ as in the Heaviside function. It is then written as \citep{Albadawi2013} 
\begin{equation}
\delta(\psi)
= \begin{cases}
0 & \text{if  $|\psi| >  \epsilon$}  \\
\frac{1}{2 \epsilon}\left[ 1+cos\left( \frac{\pi \psi}{\epsilon} \right)\right] & \text{if $|\psi| \leq \epsilon$} 
\end{cases}
\end{equation}
The value of the Dirac function is then interpolated to the face of each cell and the discretised form of $F_{\sigma}$ is finally obtained \citep{Lyras2020}. 

\subsection{Advection and correction of level set function}
In pure volume of fluid methods the interface has to be properly reconstructed allowing for the calculation of local curvature. Some of the methods widely used for the interface reconstruction are the simple line interface calculation (SLIC), PLIC and the piecewise parabolic interface calculation. The curvature has to be calculated with accuracy to avoid spurious oscillations. 
This requires the radius of curvature to remain lower than the order of the grid size (at the sub-grid scale). Instead, the method here calculates the curvature from $\psi$ which was obtained with a high-order scheme and then the interface is taken as the 0.5-isosurface without reconstructing the interface. 
For the advection of the distance function in Eq.\eqref{psiEqn} we use a third-order WENO scheme. The equation for $\psi$ can be rewritten as a hyperbolic conservation law
\begin{equation}
\frac{\partial \psi}{\partial t} + \nabla \cdot \mathbf{F}(\mathbf{u},\psi) = 0 
\label{hyperbolicPsi}
\end{equation}
where $\mathbf{F}= \psi \mathbf{u}$ is the flux-vector. Integrating over a finite volume $\Omega$ and using Gauss theorem for the volume integral of the divergence term, Eq.\eqref{hyperbolicPsi} becomes
\begin{equation}
\int_{\Omega}^{} \frac{\partial \psi}{\partial t}\text{d}\Omega + \int_{\Omega}^{} \nabla \cdot \mathbf{F}(\mathbf{u},\psi)\text{d}\Omega = \frac{\text{d}}{\text{d}t}\tilde{\psi}+\frac{1}{\Omega}\int_{a}^{b} \mathbf{F}(\mathbf{u},\psi) \cdot \mathbf{n} \partial \Omega = 0
\label{hyperbolicPsiGauss}
\end{equation} 
where $\tilde{\psi}$ is the cell-averaged value of $\psi$ considering that it does not depend on time and $\mathbf{n}$ is the normal vector of the surface $\partial \Omega$ surrounding $\Omega$. The above can be  written with respect to the surface integrals over all faces $f$ of the cell as
\begin{equation}
\frac{\text{d}}{\text{d}t}\tilde{\psi}+\frac{1}{\Omega} \sum_{f}^{}\int_{f}^{} F_{n_{f}}(\psi^-,\psi^+) \text{d} f = 0
\label{hyperbolicFinal}
\end{equation}
where the flux into the direction of the normal vector $\mathbf{n_{f}}$ of face $f$ is $F_{n_{f}}(\psi^-,\psi^+) = \mathbf{F} \cdot \mathbf{n}$. The flux is calculated considering the solutions at both cells that share the same face. The reconstructed value from the cell that the level set function is advected is denoted with $\psi^-$, and $\psi^+$ is the value of the neighbouring cell that shares $f$. The Riemann problem is solved locally for $\psi^-$ and $\psi^+$ \citep{Toro1997}. The interpolated $\psi$ to the face $f$ can be substituted with its polynomial representation leading to higher order cell-centre interpolation. 
Eq.\eqref{hyperbolicFinal} is explicitly discretisatised in time using a TVD third-order
Runge-Kutta method \citep{li2002third,gottlieb1998total}. 
The timestep in the TVD third-order Runge-Kutta discretisation has to be chosen without violating the stability of the scheme with a time-step such as $\Delta t \leq  (Co \cdot \Omega^{1/3}) / \min_f(\mathbf{u_f})$ for unstructured and structured meshes \citep{toro2005tvd}, where the Courant number $Co$ used here remained less than 0.5 for stability. 
Since the level set function is no longer a signed distance function, a re-initialisation step is made in order to satisfy the Eikonal equation. 
For this purpose the method presented in \citep{Lyras2020} is implemented. This method is second-order in space and is suitable for both structured and unstructured meshes with computational cells of arbitrary shape.  
A local search is performed for all the faces $f$ of all the cells $P$ that belong at the interface to identify which neighbouring cells $N$ that also share $f$ with $P$ also belong at the interface. The number of these cells is denoted with $N_p$. The upwind differences for these $N_p$ cells are calculated as $\partial_{f} \psi_{P}= (\psi_{P}- \psi_{N})/(x_P - x_N)$. The target value of the level set is written as 
\begin{equation}
d_{P}= \frac{\psi_{P}}{(N_{P,\Gamma}\sum_{k=1}^{N_{P,\Gamma}}(\partial_{f_k} \psi_{f_k})^2)^{1/2}}
\label{Eq:5}
\end{equation}
Next, all the cells at the interface that have negative curvature $\mathbf{\kappa}$ (calculated from the interface normal $\mathbf{n}$, $\kappa= \nabla \cdot \mathbf{n}$) or satisfy the condition $\kappa=0$ and $\tilde{\psi} \leq 0$ are considered (\citep{Hartmann2008,Lyras2020}). For the $M_P$ neighboring cells sharing the same face $f_k$ with cell $P$ but with an opposite sign for $\psi_P$, the target value of the LS function is then calculated as \citep{Hartmann2010} 
\begin{equation}
d_{P}=\psi_P\frac{\sum_{k=1}^{M_P}d_{f_{k}}}{\sum_{k=1}^{M_P}{ \psi_{f_{k}}}}
\label{Eq:7}
\end{equation}
where all the $M_{k}$ cells adjacent to $P$, sharing the face $f_k$ with a corresponding $d_{f_{k}}$ for which $\psi_P \psi_{f_{k}} \leq 0$, are considered. The value $\psi_{f_{k}}$ corresponds to the adjacent cell. The level set function in cell $P$ is calculated at the interface from $\psi_P= d_P$ \citep{Hartmann2008,Lyras2020}. The following re-initialisation equation is solved in steady state for the rest of the cells adjacent to the interface with $\alpha=$0 or 1 using a predefined time-step $\tau$
\begin{equation}
\psi_d^{n+1}=\psi_d^{n}-\Delta \tau S(\psi)( |\nabla \psi|-1)
\label{Eq:8}
\end{equation}
where $S(\psi) = \psi/\sqrt{ \psi^{2}+|\nabla \psi|^2 \Delta x^2}$. The calculated $\psi$ for the WENO scheme is used for calculating the gradient $\nabla \psi$. As an initial guess, the $\alpha$-0.5 isosurface is used as for the signed distance function, $\psi_0 = (2\alpha -1)\Delta x$. 

The gradient magnitude is calculated as $|\nabla \psi| \cong G(D^{-}_{x}\psi^n,D^{+}_{x}\psi^n,D^{-}_{y}\psi^n,D^{+}_{y}\psi^n,D^{-}_{z}\psi^n,D^{+}_{z}\psi^n)$, where $G$ is the Godunov-Hamiltonian of the level set function based on the values from the previous iteration through all faces of each cell considered. Here, the terms $D^{-}_{x}\psi^n, D^{+}_{x}\psi^n$,$D^{-}_{y}\psi^n, D^{+}_{y}\psi^n$, $D^{-}_{z}\psi^n, D^{+}_{z}\psi^n$ are the first order approximations of the gradient of $\psi$ along $x,y,z$ directions depending on whether the upwind "-" or the downwind cell is considered "+". For instance, for the x-axis, the following expressions are used
\begin{linenomath*}
\begin{align}
D^{-}_{x}\psi^n = \frac{\psi^n_{i} - \psi^n_{i-1}}{\Delta x},  &   \hspace{1cm} D^{+}_{x}\psi^n = \frac{\psi^n_{i+1} - \psi^n_{i}}{\Delta x}
\end{align}
\end{linenomath*}
The need to use first order terms arises from the large gradients across the interface that require an accurate and stable method to calculate $G$ \citep{Sethian1996,Sussman2000,Hartmann2008}. Here, the normal gradient of the level set function $\nabla^{\perp}_{f}$ is calculated for all the faces $f$ based on the orientation of the normal at the face. For instance, for the x-direction, the expression used here reads
\begin{equation}
\nabla^{\perp}_{f} = \alpha_{corr}(\psi_{P}-\psi_{N})/|\Delta x|\hat{x}
\end{equation}
where $\hat{x}$ is the unit vector at x-axis and $\alpha_{corr}$ is the inverse cosine of the angle between the cell centres and the normal face. For meshes with significant non-orthogonality, an extra term is added to take into account the angle between the cell centres at the face $f$ as in \cite{Lyras2020}. 

The Hamiltonian-Godunov term is calculated as 
\begin{equation}
G= \sqrt{max(a^{2}_{x}) +max(a^{2}_{y}) + max(a^{2}_{z})}
\end{equation}
For the unstructured meshes, the upwind-downwind cells are determined from the inner product of the interpolated gradient of $\psi$ at the face with the normal unit vector \citep{Dianat2017}. 

At x-axis, if $\psi <0$ and $\mathbf{\Delta x} \cdot \hat{x} < 0$ or $\psi >0$ and $\mathbf{\Delta x} \cdot \hat{x} > 0$ the term $a_x$ is calculated as  
\begin{equation}
a_{x} = min \left(\nabla^{\perp}_{f}\psi \cdot \hat{x} \right)
\end{equation}
If $\psi <0$ and $\mathbf{\Delta x} \cdot \hat{x} > 0$ or $\psi >0$ and $\mathbf{\Delta x} \cdot \hat{x} < 0$ then  
\begin{equation}
a_{x} = max \left(\nabla^{\perp}_{f}\psi \cdot \hat{x} \right)
\end{equation}    
The steady solutions of Eq.~\eqref{Eq:8} are distance functions and $\psi_d (\mathbf{x},\tau)$ has the same zero-level as $\psi$.

The timestep for the steady state iterations $\Delta \tau$, is selected so that an accurate value of the level set is re-initialised within a small number of iterations. $\Delta \tau$ is a percentage $\Delta x$ \citep{Prosperetti2009}. For the test in this study, $\Delta \tau= 0.14(\Delta x \Delta y \Delta z)^{\frac{1}{3}}$ has been chosen. The iterations for redistancing equation, also depends on the interface thickness $2\epsilon \Delta x$ \citep{Prosperetti2009} and $\epsilon=1.6$ for the tests here. 

Once $\psi$ is re-initialised applying the re-initialisation procedure, to obtain the signed distance function, and the interface at the boundaries is corrected the new interface curvature is calculated. The properties of the mixture are updated using the level set function. For instance, density and viscosity are calculated as follow
\begin{equation}
\rho = \rho_{1}H(\psi)+\rho_{2}(1-H(\psi))
\end{equation} 
\begin{equation}
\mu = \mu_{1}H(\psi)+\mu_{2}(1-H(\psi))
\end{equation}  

\begin{figure}[H]
    \vspace{6pt}
    \centering
    \includegraphics[scale=0.6]{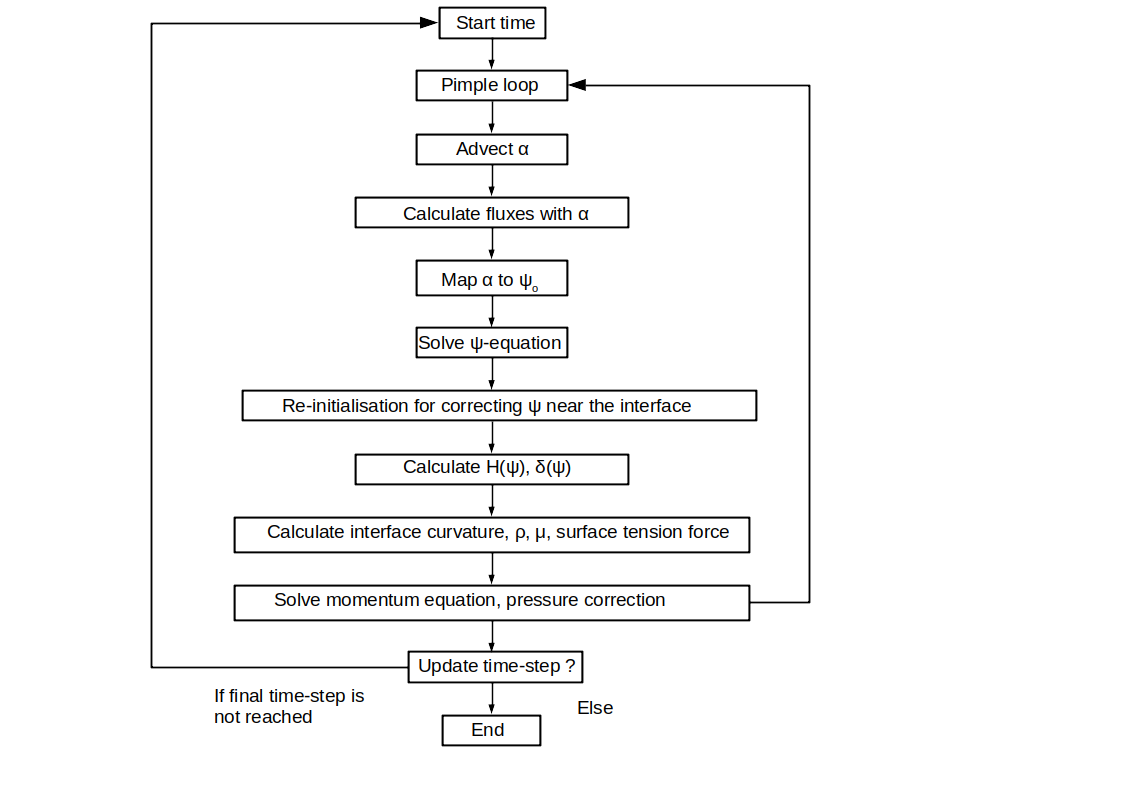}
       \centering
    \caption{Summary of the numerical algorithm used here for solving the equations. }
    \label{fig:algorithm_chart}
\end{figure}

The velocity and pressure are calculated from the Navier-Stokes equations using the pressure implicit with splitting of operators (PISO) algorithm \citep{Issa1986}. Once the pressure in its discretised form is solved, the new pressure is used for updating the fluxes within the timestep. In the momentum equation, the pressure is updated by reconstructing the face-based pressure gradients into a cell-centered gradient.
A summary of the numerical algorithm is shown in Fig.~\ref{fig:algorithm_chart}

\section{Numerical tests and discussions}
The developed method here was tested for evaluating its accuracy for various two-phase flows scenarios. The tests include fluids with different properties (density and viscosity) in two and three dimensions. First the rotating sphere test is presented wherein the ability of the method to accurately resolve thin rotating filaments which undergo a strong deformation. The dam break test is presented next, in which the obtained results of the new method are compared against experimental results. In all tests both structured and unstructured meshes were used and the results are compared with the results obtained in other numerical studies for different mesh resolutions. The application of the method with the ELSA approach is presented illustrating the capabilities of the developed solver to simulate liquid jet atomisation. 

\subsection{Rotating disc in a non-uniform velocity field}
This test case proposed by \cite{LeVeque1996} is used for testing the capability of the methodology to capture the interface of a two-dimensional rotating disc with a time-dependent interface deformation. A circular liquid disc ($l$), that is initially still, is placed at $(x,y)=(0.5,0.75)$ of a square domain $1 \cdot 1$. The fluid outside the disc is filled with another gaseous fluid ($g$) with density and viscosity $\rho_g= 1 kg/m^3$, $\nu_g =10^{-5} m^2/s$. The fluid of the disc has a density and a viscosity respectively equal to $\rho_l =1000 kg/m^3$, $\nu_l =10^{-6} m^2/s$. The disc rotates under the influence of the time-varying velocity field which is given as a function of the period $T=8s$ by 
\begin{equation}
u(x,y,t)=-sin^2(\pi x)sin(2\pi y)cos\left( \frac{\pi t}{T} \right)
\label{1}
\end{equation}
\begin{equation}
v(x,y,t)=sin(2\pi x)sin^2(\pi y)cos\left( \frac{\pi t}{T} \right)
\label{2}
\end{equation}
At $t=0$, the disc starts rotating with velocity that varies at each point $(x,y)$, deforming the interface and changing the circular shape of initial radius $R=0.15$. The deformation for the disc continues until $t=T/2$ wherein velocity is zero and the flow is reversed due to Eq.~\eqref{1} and causes the disc to rotate back to its original position.
 
\begin{figure}[H]
 \vspace{6pt}
\centering
\begin{subfigure}{.35\textwidth}
  \centering
  \includegraphics[width=0.8\linewidth]{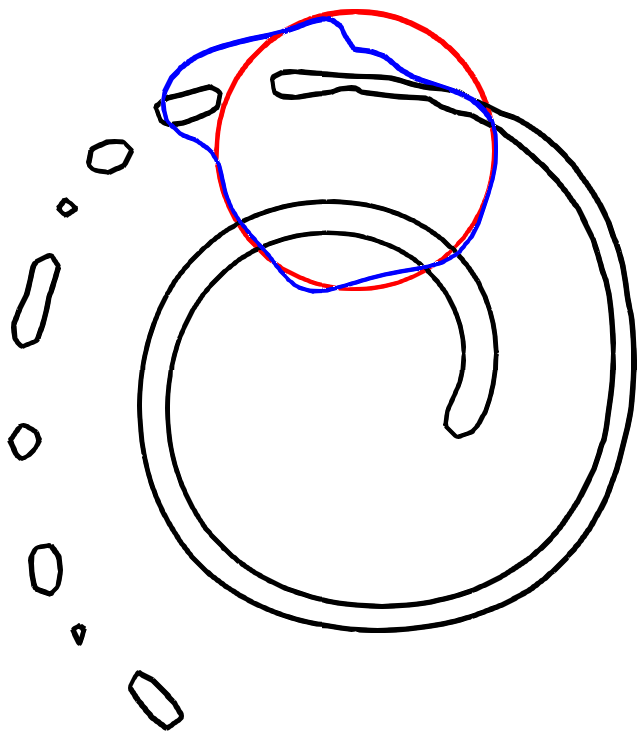}
  \caption{Coarse quadrilateral}
  \label{fig:sub1}
\end{subfigure}%
\begin{subfigure}{.35\textwidth}
  \centering
  \includegraphics[width=0.8\linewidth]{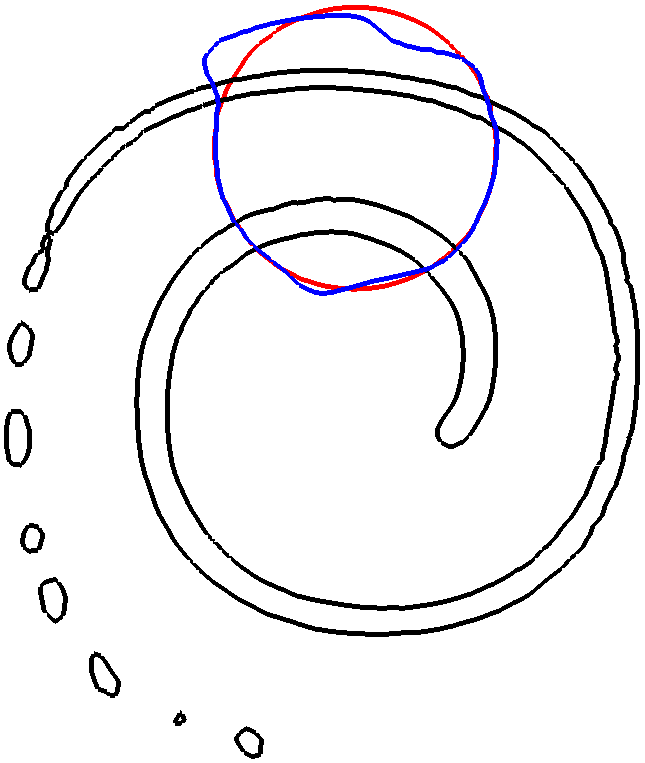}
  \caption{Coarse triangular}
  \label{fig:sub2}
\end{subfigure}

\begin{subfigure}{.35\textwidth}
  \centering
  \includegraphics[width=0.8\linewidth]{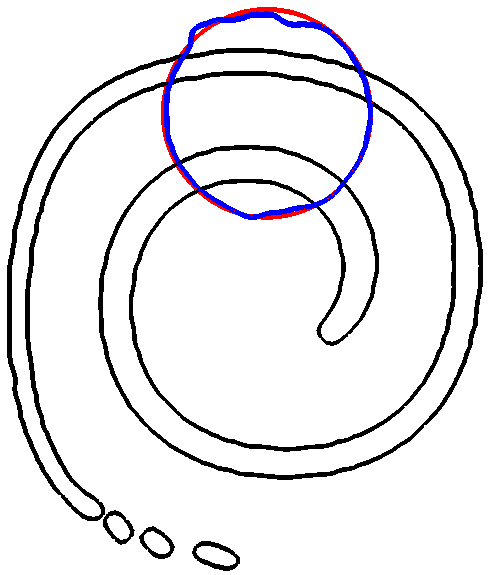}
  \caption{Medium quadrilateral}
  \label{fig:sub3}
\end{subfigure}%
\begin{subfigure}{.35\textwidth}
  \centering
  \includegraphics[width=0.8\linewidth]{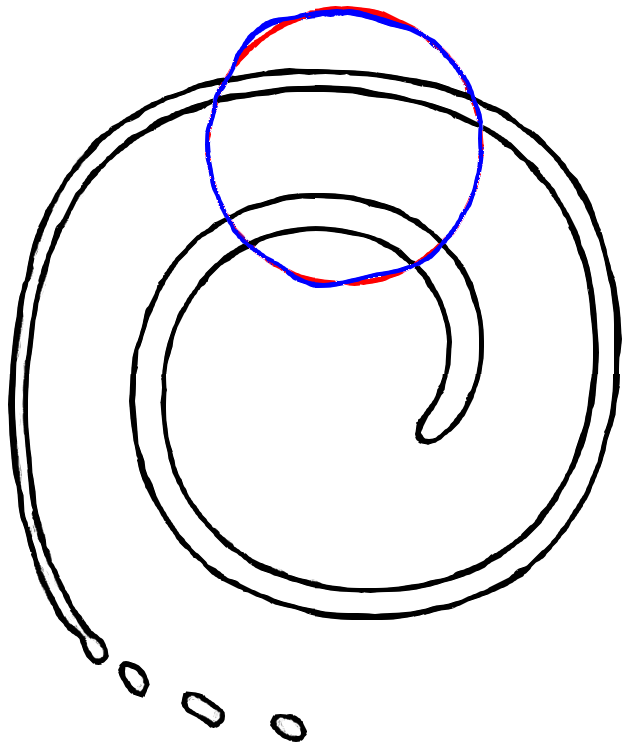}
  \caption{Medium triangular}
  \label{fig:sub4}
\end{subfigure}%

\begin{subfigure}{.35\textwidth}
  \centering
  \includegraphics[width=0.8\linewidth]{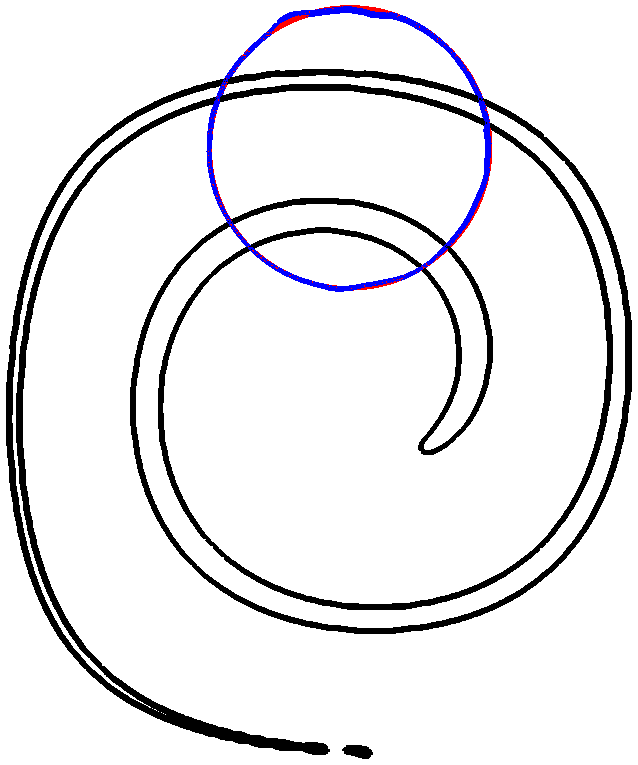}
  \caption{Fine quadrilateral}
  \label{fig:sub5}
\end{subfigure}%
\begin{subfigure}{.35\textwidth}
  \centering
  \includegraphics[width=0.8\linewidth]{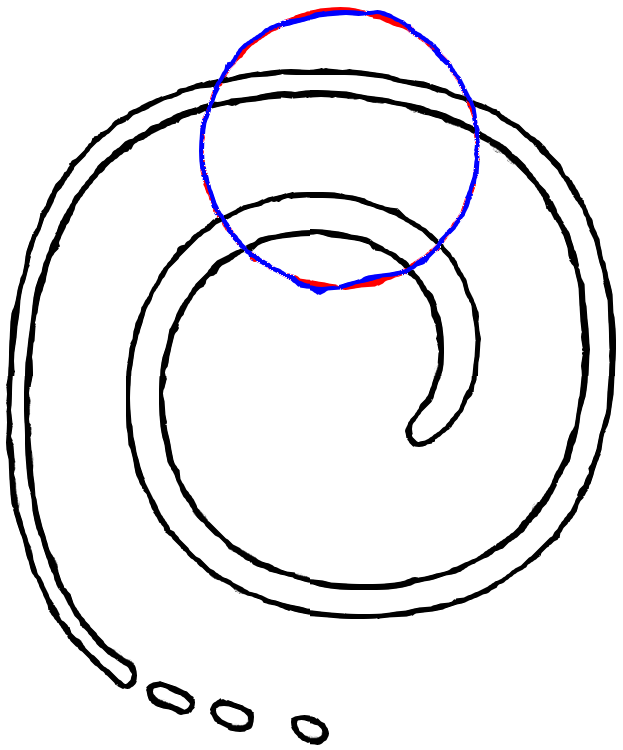}
  \caption{Fine triangular}
  \label{fig:sub6}
\end{subfigure}%
\caption{Two-dimensional rotating disc test for level set-VOF method at $t=T/2$. The initial (light purple line) at $t=0$ and final position of the zero-level set isosurface (blue line) at $t=T$ are indicated.}
\label{fig:2DrotDisk_LS_performance}
\end{figure}

The meshes used for this test were made of quadrilateral and triangular elements. Three different levels of refinement were used for each type of mesh. The mesh resolution for structured the meshes were $64 \times 64$, $128 \times 128$, $256 \times 256$ and for the unstructured meshes $4 096$, $16 016$ and $64 044$ cells. As an indicator for evaluating the capability of the method to preserve the shape of the rotating disc, we use the following function 
\begin{equation}
E_{\alpha}=\frac{\sum_{i}^{ }V_i|\alpha_i-\alpha_{exact}|}{\sum_{i}^{}V_i\alpha_{exact}}
\end{equation}
This error is defined at each timestep considering the volume $V_i$ of each cell $i$ and the volume fraction $\alpha_i$ compared to the exact solution $\alpha_{exact}$ which is the value of $\alpha$ at the initial position of the rotating disk. The results for the shape preservation error are shown in Table~\ref{table:2DRotatingDisk_Ea}.
The presented method is characterised by low shape preservation error which remains approximately of order lower than $10^{-5}$ for all the meshes. The lowest values for $E_{\alpha}$ are obtained for the finest meshes wherein the mesh resolution is adequate for capturing the thin ligament sheet that is observed during the simulation. The impact of the number of faces of the cells is also shown in Table~\ref{table:2DRotatingDisk_Ea}. The best results are achieved for the structured mesh with quadrilateral cells. Triangular cells gave higher error values that can be justified from the lower number of faces they have compared to the other mesh types. In general the methodology presented here is applicable to different types of meshes providing results with low shape preservation error.  
A more detailed description for the evolution of the rotating disc is provided in Fig.~\ref{fig:2DrotDisk_LS_performance} showing the initial positions of the disc at $t=0$ (in red colour), the final positions of the disc at $t=T$ (in blue colour) and the disc deformation at $t=T/2$ (in red colour). 
The mesh resolution used, were kept for all tests (structured and unstructured) at the similar number for each refinement level.

The grid size has a major impact on the VOF advection and the level set equation Eq.\eqref{psiEqn}. 
In all cases $E_{\alpha}$ decreases with higher mesh resolution, with the increased sharpness of the method giving more accurate surface flux and curvature calculations.
This is more evident at the maximum streching ($t=T/2$) where the tail of the disc becomes very thin and is comparable to the cell size. For the coarse meshes, resolving the ligament stretching is not possible with the ligament fragmenting into droplets. When the flow is reversed the fragmentation remains which influences  $E_{\alpha}$. 

The presented methodology is compared to other relevant numerical works for the rotating disc test case and the results are shown in Table~\ref{table:2DRotatingDisk_L1}. In the comparisons the $L_{1}(\alpha)$ error norm for the volume fraction is used.
The accuracy of the obtained results with the method presented here is similar or lower to other works regarding this test. The obtained error $L_{1}(\alpha)$ is lower than the results in ~\cite{Aulisa2003} that used PLIC and achieves similar results with the THINC scheme (tangent of hyperbola for interface capturing) in ~\cite{Yokoi2007,Xie2017,Xiao2011}. Compared to other coupled LS methods such as the level set-moment of fluid and level set-vof methods in ~\cite{Jemison2013} the $L_{1}(\alpha)$ error is similar or smaller for the different grid resolutions. The results obtained here are close to the results published by \cite{Qian2018} with the coupled level set method using the THINC scheme where the interface is represented with polynomials from the level set, rather than using a plane and quadratic surface representation. For the comparisons we use values $L_1$ for two cases: one for a polynomial order 1, denoted with THINC-LS (P1) and one with polynomial order 4, denoted with THINC-LS (P4). The results obtained here were between the two methods for P1 and P4 and had generally smaller error than the first-order polynomial representation, being closer to the P4 results compared to the P1 case. 

\begin{table}
\caption{$E_{\alpha}$ calculated using different meshes for the 2D rotating disc case.}
\centering
\label{table:2DRotatingDisk_Ea}
\begin{tabular}{l l l}
\hline
Mesh & Resolution & $E_{\alpha}$ \\
\hline
 & $64^2$ & -1.12$\cdot 10^{-7}$\\

Structured & $128^2$ & -4.58$\cdot 10^{-8}$\\

 & $256^2$  & -9.87$\cdot 10^{-9}$\\
 
  \midrule
 
 & $4096$ & -5.77$\cdot 10^{-6}$\\

Unstructured &$16016$ & 1.14$\cdot 10^{-6}$\\

 &$64044$ & -6.2$\cdot 10^{-7}$\\
 
\hline
\end{tabular}
\end{table}

\begin{table}
\caption{Comparisons of the methods using quadrilateral meshes for the two-dimensional rotating disc case. The first order norm $L_{1}(\alpha)$ is calculated for the three different meshes.}
\centering
\label{table:2DRotatingDisk_L1}
\begin{tabular}{l l l l}
\hline
Authors / Mesh resolution & $32^{2}$ & $64^{2}$ & $128^{2}$ \\
\hline
Rider–Kothe/Puckett ~\citep{Rider1998} & 4.78$\times 10^{-2}$ & 6.96$\times 10^{-3}$ &  1.44$\times 10^{-3}$ \\

DS-CLSMOF ~\citep{Jemison2013} & 2.92$\times 10^{-2}$ & 5.51$\times 10^{-3}$ & 1.37$\times 10^{-3}$  \\

DS-CLSVOF ~\citep{Jemison2013} & 5.45$\times 10^{-2}$ & 1.05$\times 10^{-2}$ & 1.74$\times 10^{-3}$  \\

PLIC ~\citep{Aulisa2003} & 2.53$\times 10^{-2}$ & 2.78$\times 10^{-3}$ & 4.8$\times 10^{-4}$  \\

Markers-VOF ~\citep{Lopez2005} & 7.41$\times 10^{-3}$ & 2.78$\times 10^{-3}$ & 4.78$\times 10^{-4}$ \\

THINC/WLIC ~\citep{Yokoi2007} & 4.16$\times 10^{-2}$ & 1.61$\times 10^{-2}$ & 3.56$\times 10^{-3}$  \\

THINC/QQ ~\citep{Xie2017} & 6.70$\times 10^{-2}$ & 1.52$\times 10^{-2}$ & 3.06$\times 10^{-3}$  \\

THINC/SW ~\citep{Xiao2011} & 3.90$\times 10^{-2}$ & 1.52$\times 10^{-2}$ & 3.96$\times10^{-3}$  \\

THINC-LS (P1) ~\citep{Qian2018} & 6.71$\times 10^{-2}$ & 1.53$\times 10^{-2}$ & 2.27$\times 10^{-3}$  \\

THINC-LS (P4) ~\citep{Qian2018} & 2.85$\times 10^{-2}$ & 3.39$\times 10^{-3}$ & 6.79$\times 10^{-4}$  \\

Presented method & 3.14$\times 10^{-2}$ & 4.62$\times 10^{-3}$ & 9.12$\times 10^{-4}$  \\

\hline
\end{tabular}
\end{table}

\subsection{Rotating sphere}
In this three-dimensional test case, a sphere of liquid initially at rest starts to rotate under the influence of a time-dependent velocity field \citep{LeVeque1996}. The initial sphere has a radius such as $R=0.15m$ and is placed inside a box $[0,1]^3$ with its centre at $(0.35, 0.35, 0.35)$. The velocity field is defined as
\begin{linenomath*} 
\begin{align}
\begin{split}
u(x,y,z,t)= 2sin^2(\pi x)sin(2\pi y)sin(2\pi z)cos\left(\frac{\pi t}{T}\right) \\
v(x,y,z,t)= -sin(2\pi x)sin^2(\pi y)sin(2\pi z)cos\left(\frac{\pi t}{T}\right) \\
w(x,y,z,t)= -sin(2\pi x)sin(2\pi y)sin^2(\pi z)cos\left(\frac{\pi t}{T}\right)
\end{split}
\end{align} 
\end{linenomath*} 
The period is $T=3s$ and the density and viscosity of both fluids in the test are the same as in the rotating disc case. This test case is used to assess the capability of the methodology to capture significantly distorted interface \citep{Menard2007,Albadawi2013,Roenby2016,Dianat2017}. The sphere starts deforming at $t=0$ due to the velociy difference at the interface. The rotation reverses at $t=T/2$ causing the sphere to return back to its original position $t=T$. The presented method was tested for  different mesh resolution. Both structured meshes with $32^3, 64^3, 128^3$ hexahedral elements and unstructured meshes with 81 008, 224 622, 998 810 tetrahedral elements were used. The results for the error in shape preservation $E_{\alpha}$ are shown in Table~\ref{table:3DRotatingSphere_Ea}. $E_{\alpha}$ decreases with the level set implementation. 

\begin{figure}[H]
 \vspace{1pt}
\centering
\begin{subfigure}[b]{.22\textwidth}
  \centering
  \includegraphics[width=1\linewidth]{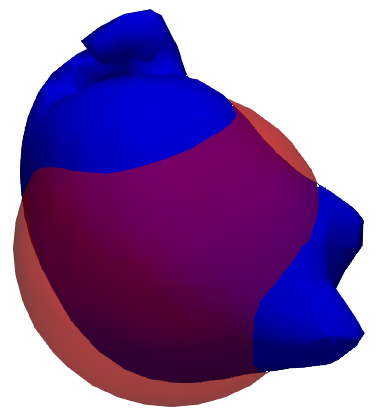}
  \caption{$32^{3}$}
  \label{fig:sub4}
\end{subfigure}%
\begin{subfigure}[b]{.24\textwidth}
  \centering
  \includegraphics[width=1\linewidth]{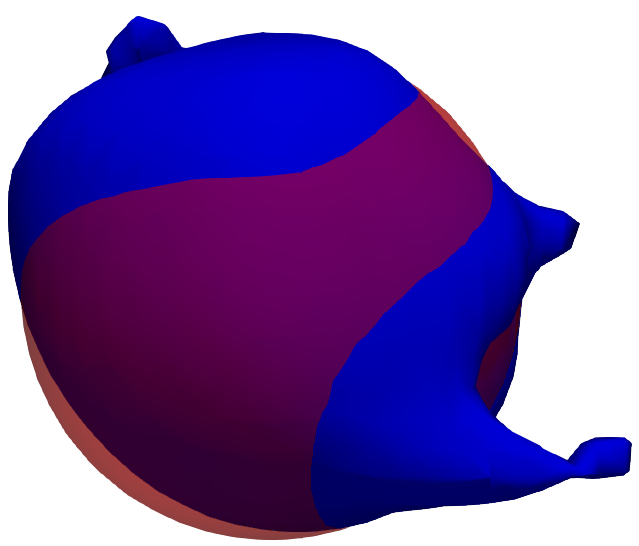}
  \caption{$64^{3}$}
  \label{fig:sub5}
\end{subfigure}
\begin{subfigure}[b]{.236\textwidth}
  \centering
  \includegraphics[width=1\linewidth]{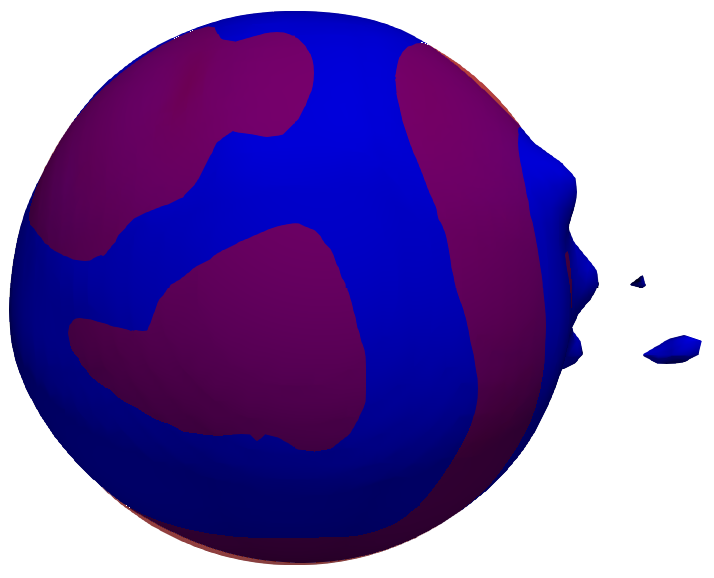}
  \caption{$128^{3}$}
  \label{fig:sub6}
\end{subfigure}%

\begin{subfigure}[b]{.26\textwidth}
  \centering
  \includegraphics[width=1\linewidth]{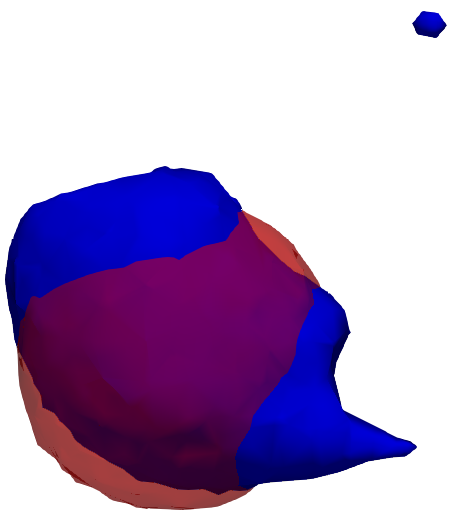}
  \caption{81008}
  \label{fig:sub4}
\end{subfigure}%
\begin{subfigure}[b]{.28\textwidth}
  \centering
  \includegraphics[width=1\linewidth]{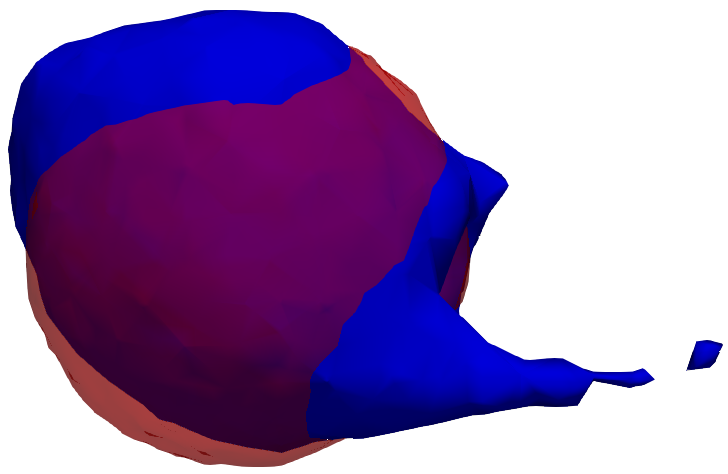}
  \caption{224622}
  \label{fig:sub5}
\end{subfigure}
\begin{subfigure}[b]{.22\textwidth}
  \centering
  \includegraphics[width=1\linewidth]{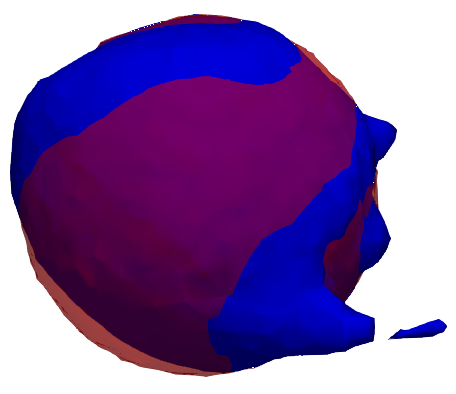}
  \caption{998810}
  \label{fig:sub6}
\end{subfigure}%

\caption{Isosurfaces for the rotating sphere in a non-uniform flow test. The 0.5-isosurface obtained with the presented method at $t=0$ (red colour) and at $t=T$ (blue colour) are shown for the different mesh resolutions.}
\label{fig:3DrotSphere_0s_3s_comparison}
\end{figure}

For lower grid resolutions the interface is under-resolved with the non-uniform flow deforming the sphere which breaks up before the flow reverses at $t=T/2$. This can also be observed in Fig.~\ref{fig:3DrotSphere_0s_3s_comparison} which provides the initial (in red colour) and final (in blue colour) shapes of the sphere for the different meshes used.  
The inadequate grid resolution also causes distortion in the sphere which results to holes that disappear with increasing the mesh resolution. This is also observed by other similar numerical works in \cite{Dianat2017,Hernandez2008,Xiao2011,Deshpande2012,Roenby2016}. This reflects on the $E_{\alpha}$ error which decreases with increasing mesh resolution. 
The error in shape preservation is lower in structured meshes, being an order of magnitude less than unstructured meshes. The presented method shows adequate accuracy levels when using tetrahedral meshes. The signed distance function correction using a targeted initialisation step for advecting level set as the one in \cite{Lyras2020} aims at improving the interface capturing and errors induced with the VOF advection.  
Comparing the behavior of $E_{\alpha}$ in the rotating disc and the error in the three-dimensional test of the rotating sphere, the latter gives higher errors most likely due to the significant distortion of the interface in the 3D test. 
In the finer mesh, only a small part of the rotating sphere is under-resolved.
The error $L_{1}(\alpha)$ for the volume fraction is shown in Table~\ref{table:3DRotatingDisk_L1} and is compared to other numerical works. $L_{1}(\alpha)$ is calculated for all cells $k$ considering their volume ($L_1$ error norm \cite{Xiao2011,Deshpande2012}) and is defined as
\begin{equation}
L_1(\alpha) = \sum_{k}(\alpha_{k}-\alpha_{exact})V_{k}
\end{equation}
where $V_{k}$ is the cell volume. The presented methodology achieved lower errors compared to other works that use pure VOF such as interFoam using MULES limiter \citep{Deshpande2012} and the THINC/SW scheme in \cite{Xiao2011}. 
Results are reasonably low and similar to the ones obtained with the Piecewise-Constant Flux Surface Calculation (PCFSC) method in \cite{liovic2006} which used a method for approximating sphere interface as piecewise planar, advecting volume in a single unsplit step using multi-dimensional fluxes. 
In the PLIC methods mentioned in Table~\ref{table:3DRotatingDisk_L1}, the sphere break up might be attributed to having only one piecewise linear segment in the cut cell. Hence, when the interface filament is smaller than the grid size, more than one piecewise linear segments are required for describing the shape of the sphere. However, the PCFSC method was applied for orthogonal meshes. The presented methodology here is applied to arbitrary meshes unstructured or polyhedral offering similar accuracy for calculating volume fraction. 
The results for THINC-LS (P1) and THINC-LS (P4) in \cite{Qian2018} were very close to each other, with the results here being closer to THINC-LS (P1) and slightly higher. 

\begin{figure}[H]
    \vspace{6pt}
    \centering
    \includegraphics[scale=0.16]{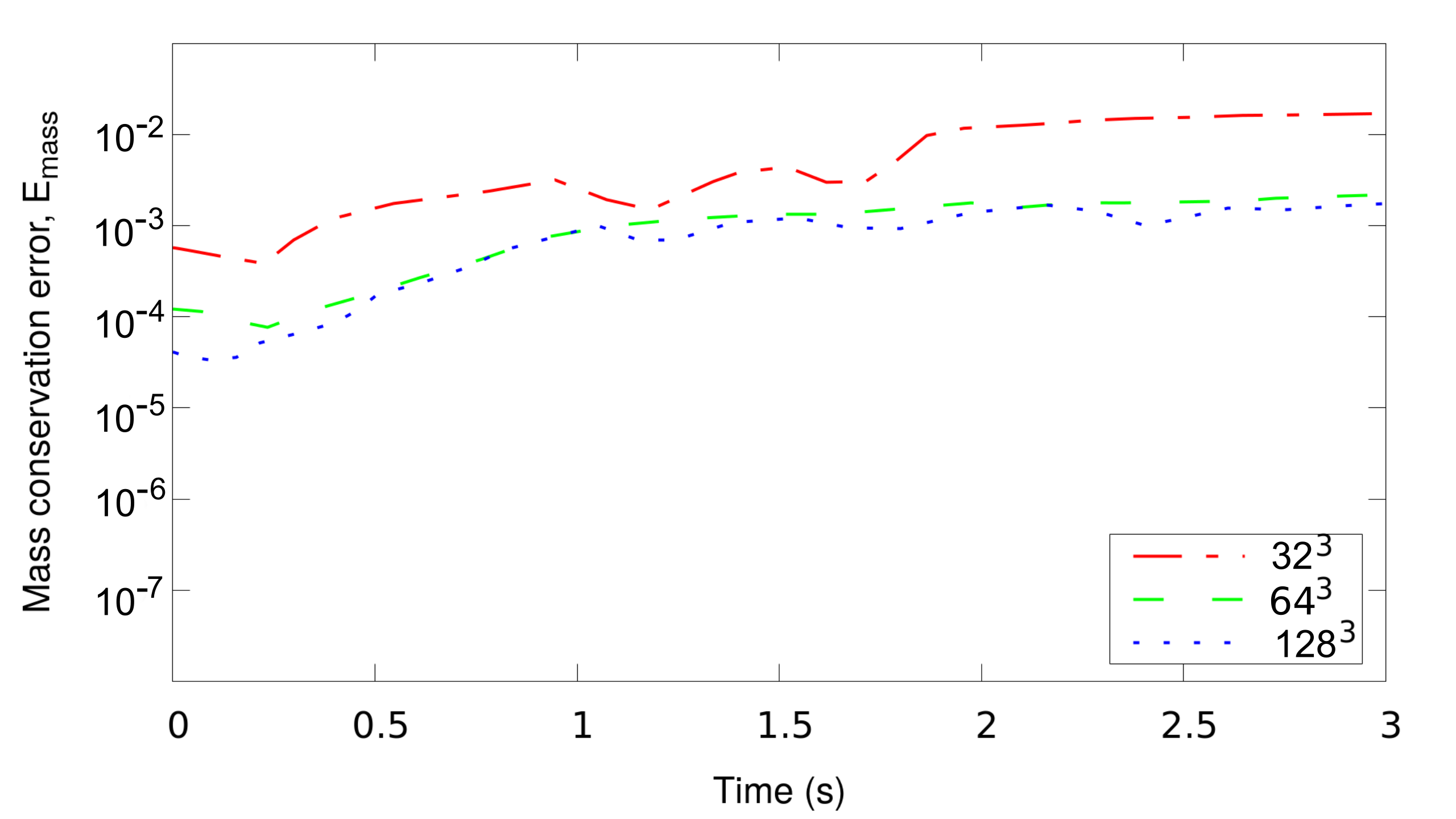}
       \centering
    \caption{Mass conservation error for the three-dimensional rotating sphere test for the three grids. }
    \label{fig:3D_rotating_sphere_mass_conservation}
\end{figure}

The mass conservation error $E_{mass}$ evolution is shown in Fig.~\ref{fig:3D_rotating_sphere_mass_conservation} for the three meshes used in Table~\ref{table:3DRotatingDisk_L1}. The presented method has showed generally reasonable mass conservation error for the rotating sphere, with a small mass loss lower than $10^{-2}$.
     
In terms of simulation time, the presented method in general was found to be reasonably slower than for instance in the case of advecting only the volume fraction $\alpha$. However, in some of our tests we have seen times that varied: The method here was generally from 35$\%$ slower up to 30$\%$ faster than the standalone VOF method as shown in Table~\ref{table:simulationTimes}. The latter behaviour was observed for finer meshes for some cases such as the rotating sphere. Consequently, this indicates that the overhead created with the coupled method might vary depending on the problem. Although deeper analysis is required to study the parameters that influence the simulation time, this was attributed to the better pressure-velocity coupling achieved when level set is used, indicated by the lower residuals for the pressure solution. For instance, for the finest mesh for the 3D rotating sphere, almost 50$\%$ smaller values were obtained compared to the ones taken with the VOF-only method.           
\begin{table}
\caption{$E_{\alpha}$ error calculated using different meshes for the 3D rotating sphere test.}
\centering
\label{table:3DRotatingSphere_Ea}
\begin{tabular}{l l l}
\hline
Mesh & Resolution & $E_{\alpha}$ \\
\hline
 & $32^3$ & -7.83$\cdot 10^{-6}$\\

Structured & $64^3$ & -6.6$\cdot 10^{-7}$\\

 & $128^3$  & -1.26$\cdot 10^{-7}$\\
 
  \midrule
 
 & 81008 & -6.13$\cdot 10^{-4}$\\

Unstructured & 224622 & -8.92$\times 10^{-5}$\\

 & 998810 & -4.97$\cdot 10^{-5}$\\
 
\hline
\end{tabular}
\end{table}
   
\begin{table}
\caption{Simulation time in seconds for the presented method and the standalone VOF method using different meshes for the 2D rotating disc and the 3D rotating sphere tests.}
\centering
\label{table:simulationTimes}
\begin{tabular}{l l l l}
\hline
Case & Resolution & Standalone VOF & Presented method \\
\hline
 & Coarse & 499 & 597 \\

2D rotating disc & Medium & 2335 & 3151 \\

 & Fine  & 3516 & 4043 \\
 
  \midrule
 
 & Coarse & 131 & 177\\

3D rotating sphere & Medium & 1352 & 1555 \\

 & Fine & 28768 & 21576\\
 
\hline
\end{tabular}
\end{table}
                 
\begin{table}
\caption{Calculated error norm $L_{1}(\alpha)$ for the three-dimensional rotating sphere test. Results are  compared with other numerical methods.}
\centering
\label{table:3DRotatingDisk_L1}
\begin{tabular}{l l l l}
\hline
Authors & $32^{3}$ & $64^{3}$ & $128^{3}$ \\
\hline
PCFSC VOF ~\citep{liovic2006} & 7.86$\times10^{-3}$ & 2.91$\times10^{-3}$ & 7.36$\times 10^{-4}$  \\

THINC/SW scheme ~\citep{Xiao2011} & 8.39$\times 10^{-3}$ & 3.47$\times 10^{-3}$ & 1.08$\times10^{-3}$  \\

THINC-LS (P1) ~\citep{Qian2018} & 7.18$\times 10^{-3}$ & 2.34$\times 10^{-3}$ & 6.14$\times 10^{-4}$  \\

THINC-LS (P4) ~\citep{Qian2018} & 5.54$\times 10^{-3}$ & 1.57$\times 10^{-3}$ & 3.79$\times 10^{-4}$  \\

interFoam ~\citep{Deshpande2012} & 9.95$\times 10^{-3}$ & 4.78$\times 10^{-3}$ & 2.03$\times 10^{-3}$  \\
                    
Present method & 8.64$\times 10^{-3}$ & 3.24$\times 10^{-3}$ & 7.04$\times 10^{-4}$  \\

\hline
\end{tabular}
\end{table}

\subsection{Three-dimensional dam break}
The dam break problem is used for the validation of the presented methodology and is used in various interface capturing methods for assessing their capability of the method to model free surface problems. The set-up of the test case has simple geometry and initial conditions and experiment data and other numerical results are available for comparisons. 
The problem consists of a three-dimensional rectangular domain of dimensions $4a \times 2.4a \times a$ with a liquid column with dimensions $2.4a \times a$ initially. The problem is dimensionalised with $a$= 0.146$m$ according to \citep{Martin1952,Koshizuka1996}. At $t=0$, the liquid column which is filled with water starts to collapse. The density $\rho_{l}$ and kinematic viscosity $\nu_{l}$ of water are such as $\rho_{l} =$1000$kg/m^3$ and $\nu_{l}=10^{-6}m^2/s$. The rest of the domain is filled with air. The air density $\rho_{g}$ and kinematic viscosity $\nu_{g}$ are respectively taken equal to $1kg/m^3$ and $10^{-4}m^2/s$.   

The initial velocity is zero and the pressure is equal to the hydrostatic pressure. Free slip boundary conditions are imposed for all the boundaries of the domain with zero normal velocity and zero tangential traction. For the top boundary of the rectangular domain the tangential velocity and normal traction are zero. The time evolution of the liquid/gas interface displacement is calculated using three different grid resolutions as in \citep{Elias2007,Zhao2014} i.e. $40 \times 10 \times 20$, $80 \times 20 \times 40$ and $160 \times 40 \times 80$ cells. 

\begin{figure}[!tbp]
  \begin{subfigure}[b]{0.62\textwidth}
    \includegraphics[width=\textwidth]{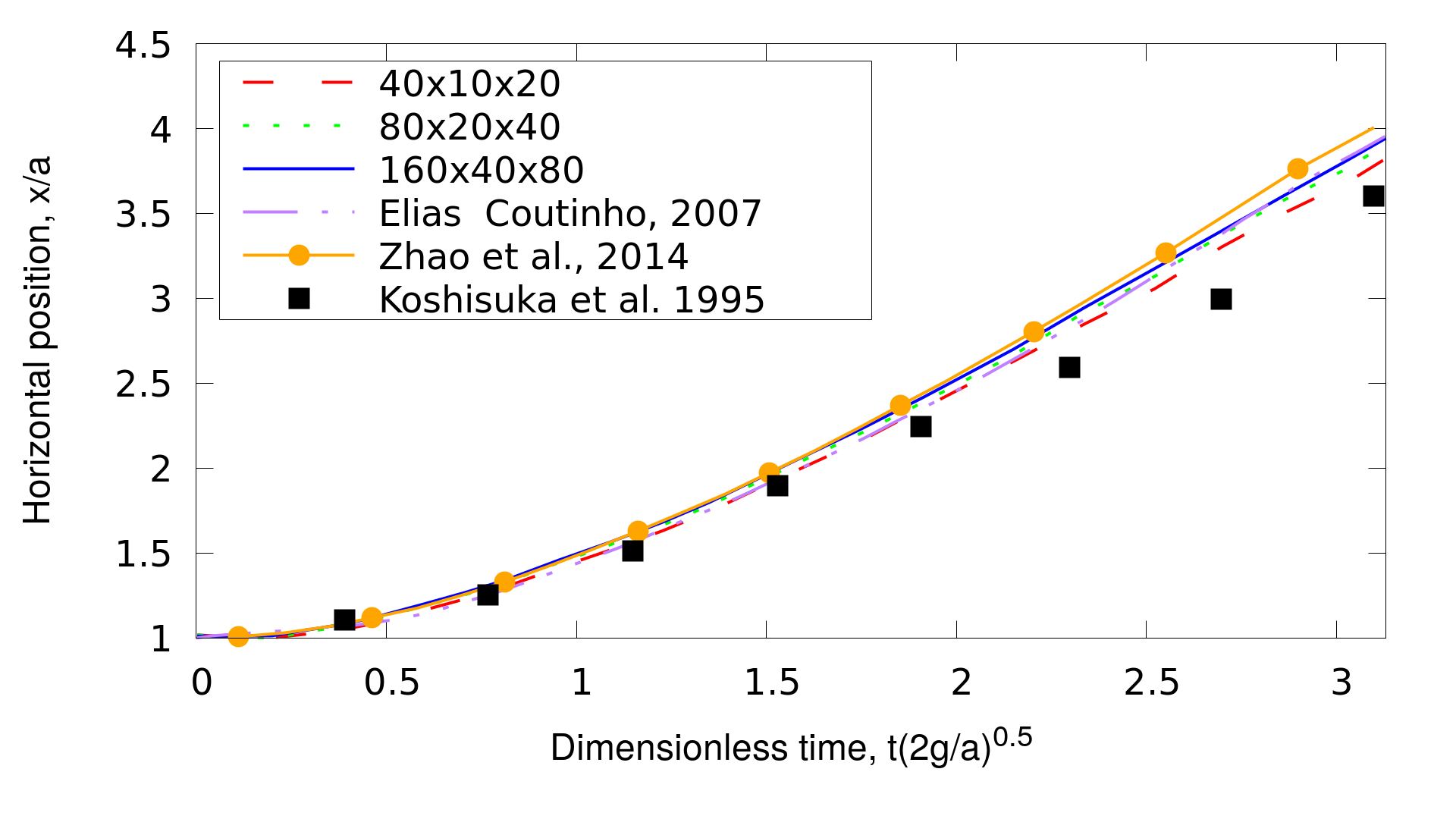}
    \caption{X-axis.}
    \label{fig:sub1}
  \end{subfigure}
  \hfill
  \begin{subfigure}[b]{0.62\textwidth}
    \includegraphics[width=\textwidth]{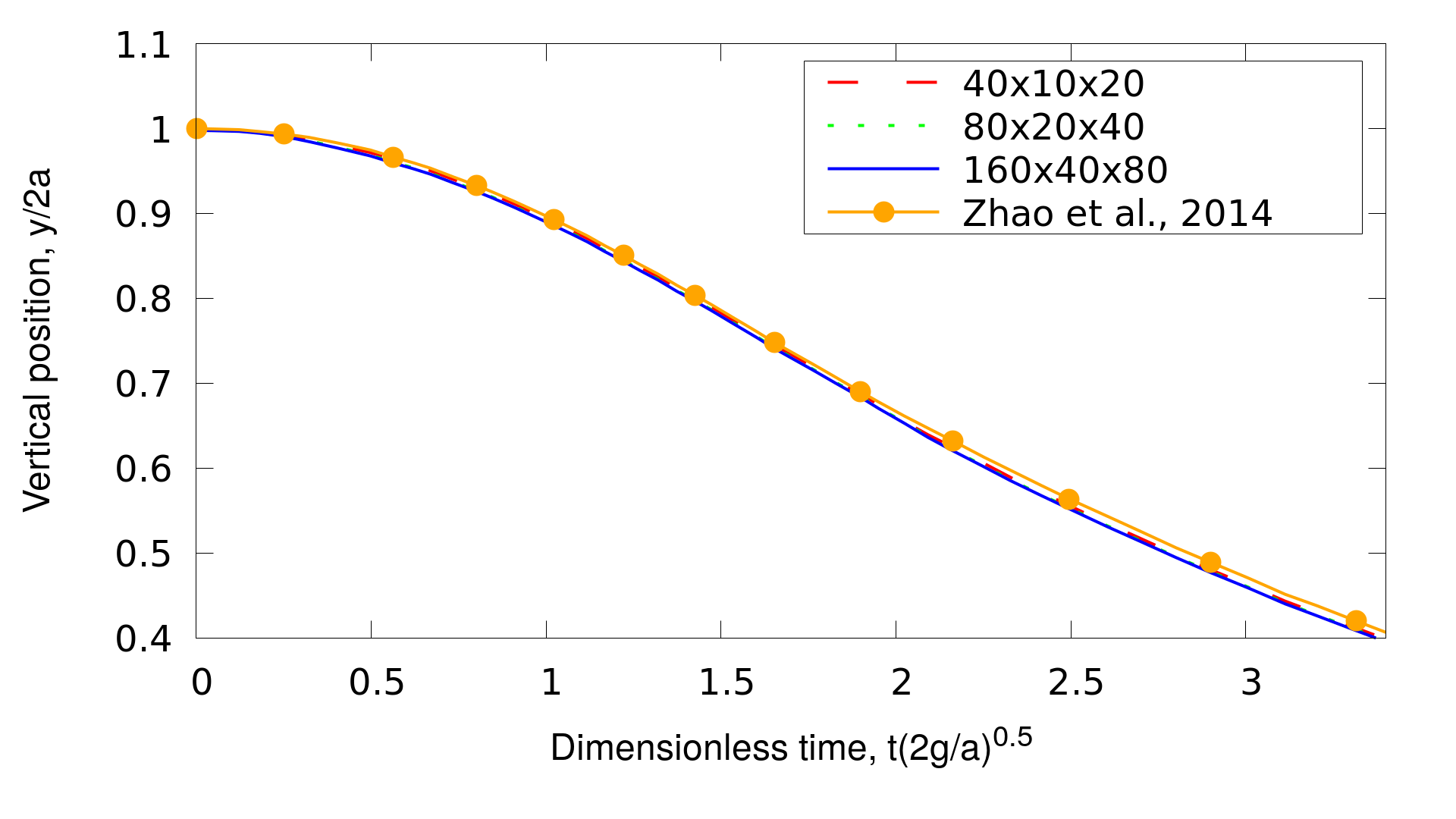}
    \caption{Y-Axis.}
    \label{fig:sub2}
  \end{subfigure}
\caption{Liquid/gas interface position along the x-axis and y-axis for the three-dimensional dam break simulation.}
\label{fig:damBreakResults_xy}
\end{figure}

The results are compared with the experimental data available in \cite{Koshizuka1996} and previous numerical studies using the LS method from \cite{Zhao2014} which is a conservative level set method based on the the re-initialisation step that consists of solving for the steady state of the regularised non-linear equation as proposed by \cite{Olsson2005}. The re-initialisation step of the particular formulation in \cite{Zhao2014} consists of a compression term for the interface profile and a diffusive flux term. The results in \cite{Olsson2005,Zhao2014} have shown that the VOF-LS coupling based on that re-initialisation step is conservative and has shown excellent results for both structured and unstructured meshes in two-dimensional and three-dimensional tests with second order accuracy.   
The calculated position of the interface both normalised with the length parameter $a$ along the horizontal (x-axis) and the vertical (y-axis) directions are shown in Fig.~\ref{fig:damBreakResults_xy}. The results are plotted against the dimensionless time $t^{*}$ $=t\sqrt{(2g/a)}$. The horizontal interface displacement predictions for the leading-edge position are in close agreement with the experimental data for the $t^{*}$ such that the leading edge reaches the wall, $x/a=4$. The accuracy in the predictions for the horizontal direction is closer initially and reasonably close to the experiment. Due to wall friction the interface motion is slowed down \citep{Koshizuka1996} which also causes the calculated interface to differ from the experiment.
Since the thickness of the boundary layer can be much smaller than the scale of the problem here and a near-wall discretisation for capturing the boundary layer can be computationally too expensive. To avoid this issue, a free-slip boundary condition is considered rather than a no-slip condition \citep{Cerquaglia2017}. In general, both boundary conditions can be employed for the dam break problem \citep{Kees2011,Elias2007}. 
The liquid front in the present simulations here propagates at similar rate than in \cite{Elias2007} although the results in \cite{Elias2007} were obtained for slightly different physical properties (the viscosity ratio in \cite{Elias2007} between the two fluids was 58 and here it was 100). All three meshes gave results in the present study reasonably close to the results reported in \cite{Koshizuka1996}.  

\begin{figure}[H]
 \vspace{6pt}
\centering
\begin{subfigure}{.47\textwidth}
  \centering
  \includegraphics[width=0.8\linewidth]{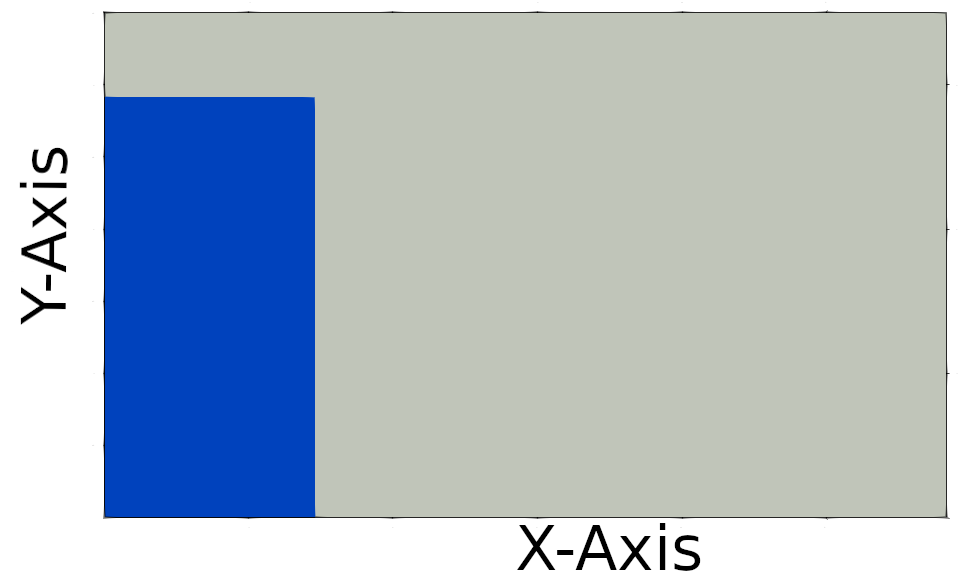}
  \caption{t=0s}
  \label{fig:sub4}
\end{subfigure}%
\begin{subfigure}{.47\textwidth}
  \centering
  \includegraphics[width=0.8\linewidth]{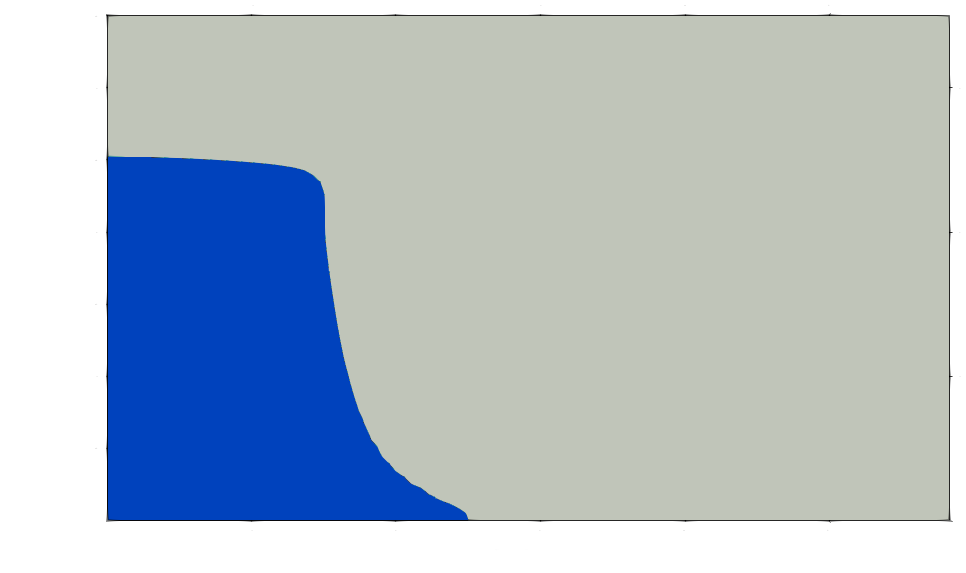}
  \caption{t=0.1s}
  \label{fig:sub5}
\end{subfigure}
\begin{subfigure}{.47\textwidth}
  \centering
  \includegraphics[width=0.8\linewidth]{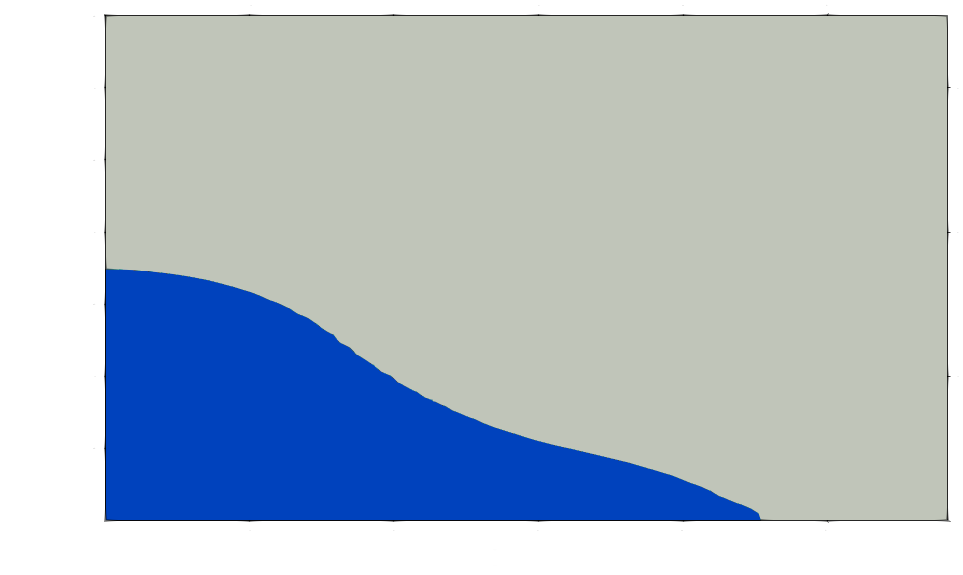}
  \caption{t=0.2s}
  \label{fig:sub6}
\end{subfigure}%
\begin{subfigure}{.47\textwidth}
  \centering
  \includegraphics[width=0.8\linewidth]{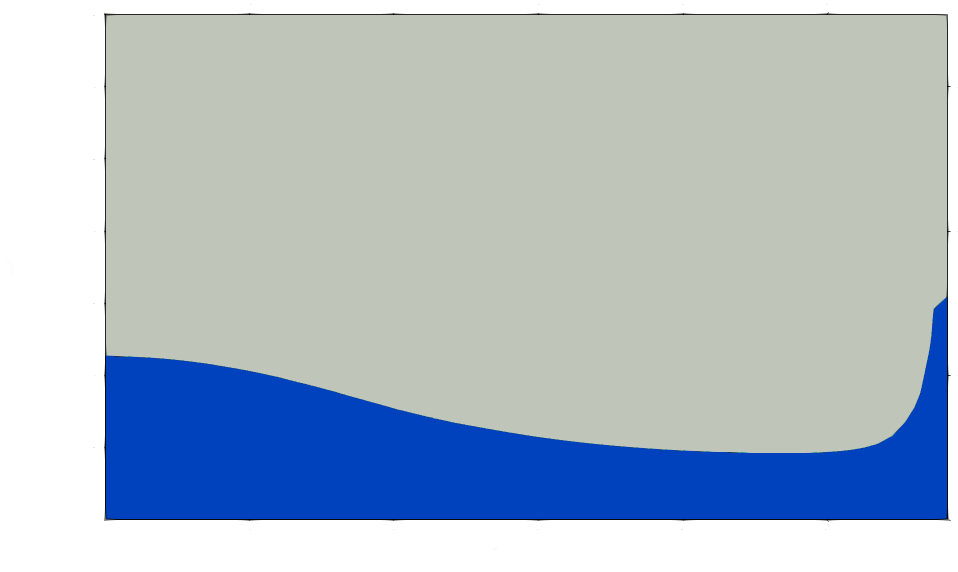}
  \caption{t=0.3s}
  \label{fig:sub6}
\end{subfigure}%

\begin{subfigure}{.47\textwidth}
  \centering
  \includegraphics[width=0.8\linewidth]{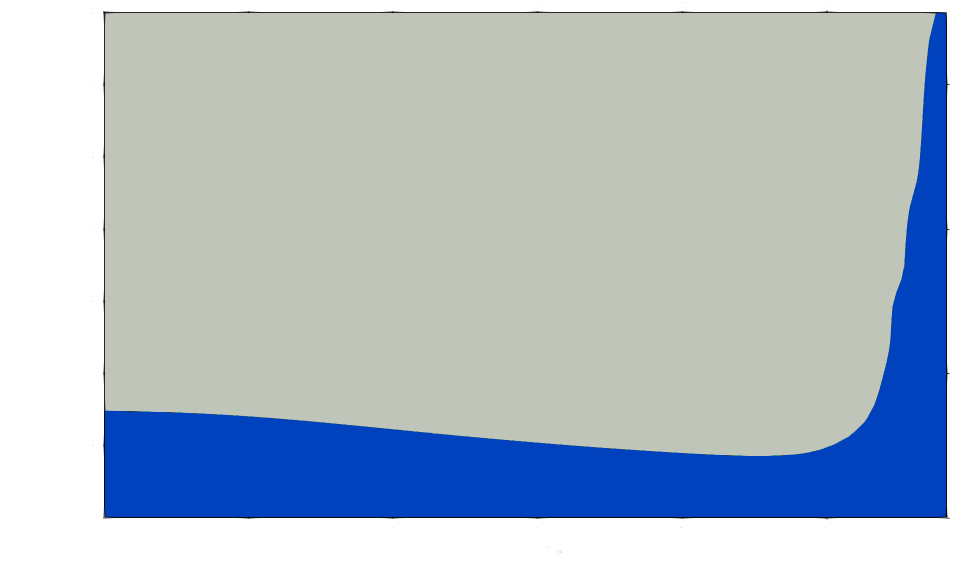}
  \caption{t=0.4s}
  \label{fig:sub6}
\end{subfigure}%
\caption{0.5-isosurface snapshots at different times for the three-dimensional dam break test. The collapse of the liquid column starts at t=0s and moves towards the right wall, moves upwards until it returns back to the bottom.}
\label{fig:3DdamBreak}
\end{figure}

\begin{figure}[h]
    \vspace{6pt}
    \centering
    \includegraphics[scale=0.16]{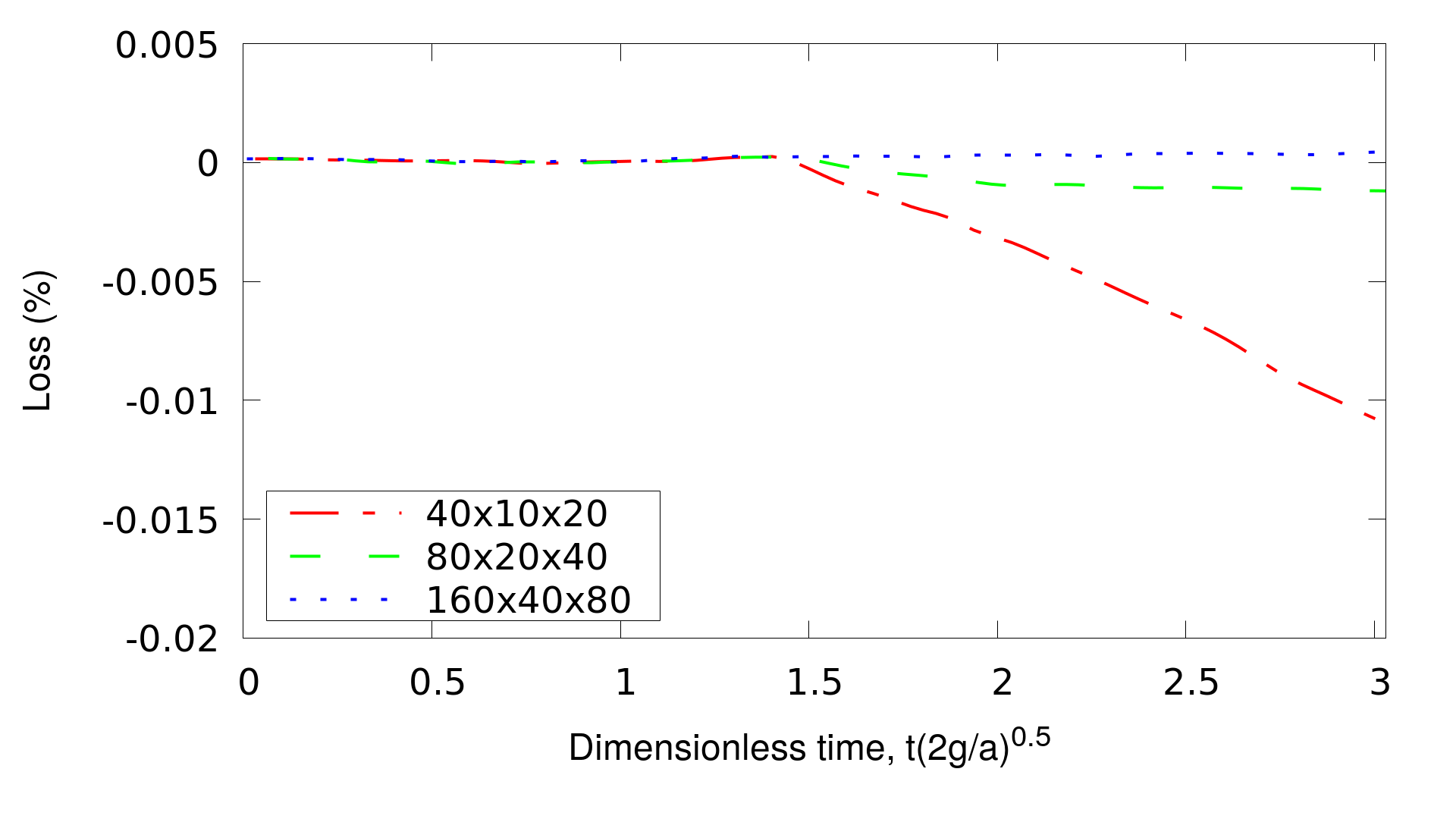}
       \centering
    \caption{Percent mass loss for the dam break case through time for different mesh resolutions.}
    \label{fig:3D_dambreak_mass_conservation}
\end{figure}

The interface displacement is shown in Fig.~\ref{fig:3DdamBreak} wherein the interface preserves its nearly flat profile for $t$ $\in [0,0.2s]$. Once the interface reaches $x/a=4$, the liquid rises upwards with the resulting formation of a layer on the wall. 
The vertical position of the interface decreases at $t^*$ $\in [0,3]$ and the present method results are very close to the results of \cite{Zhao2014} (Fig.~\ref{fig:damBreakResults_xy}b) with the same rate of change in the liquid column height for the different meshes. The time-evolution for the fluid mass loss percentage of the collapsing liquid column test is shown in Fig.~\ref{fig:3D_dambreak_mass_conservation}. The presented method has demonstrated good mass conservation for the meshes used, the mass conservation error remaining lower than $1.5 \cdot 10^{-2}$ percent. 

\subsection{Droplet impact and crown propagation}
\begin{figure}[!tbp]
  \begin{subfigure}[b]{0.42\textwidth}
    \includegraphics[width=\textwidth]{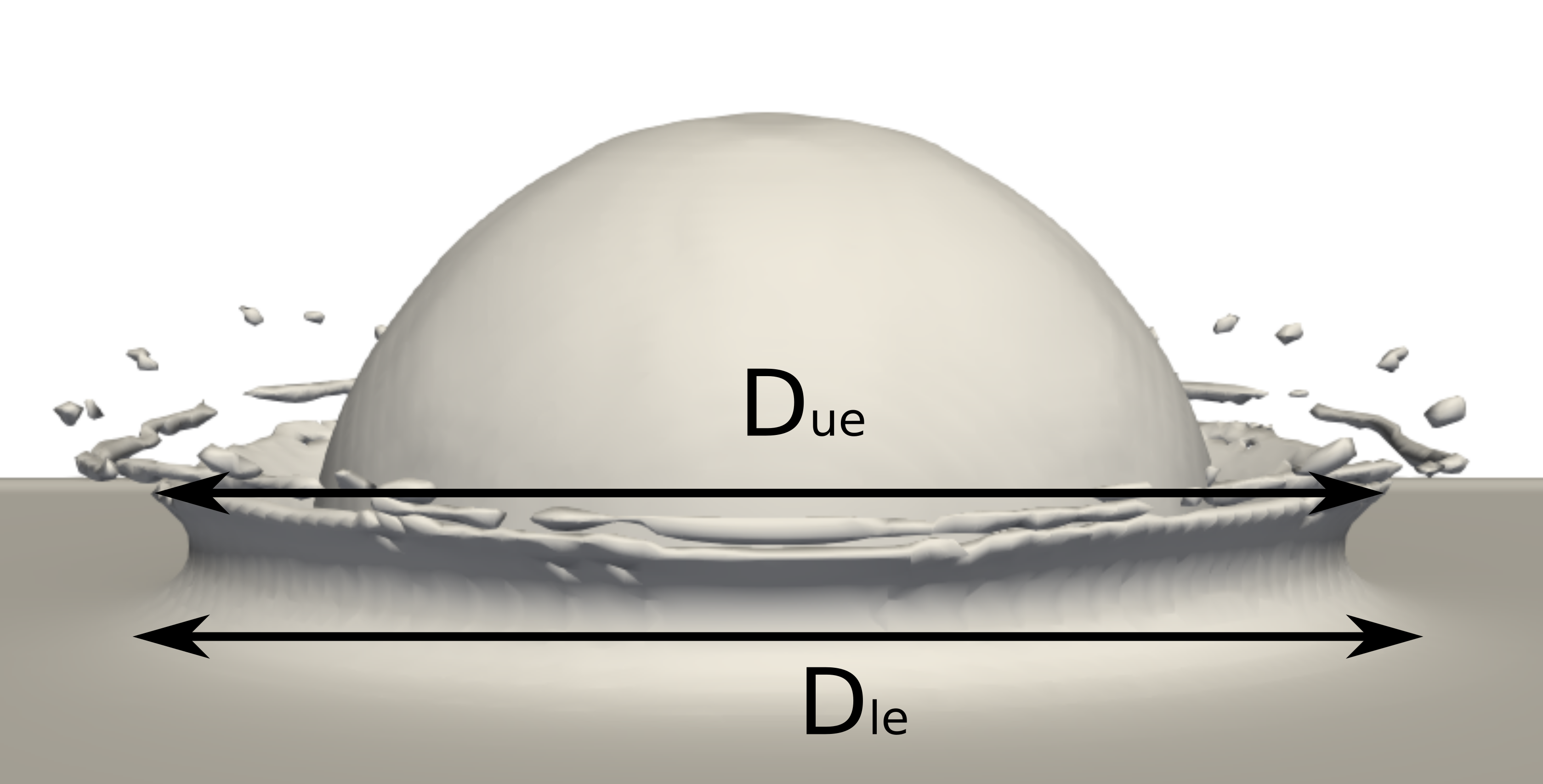}
    \label{fig:sub1}
  \end{subfigure}
  \hfill
  \begin{subfigure}[b]{0.68\textwidth}
    \includegraphics[width=\textwidth]{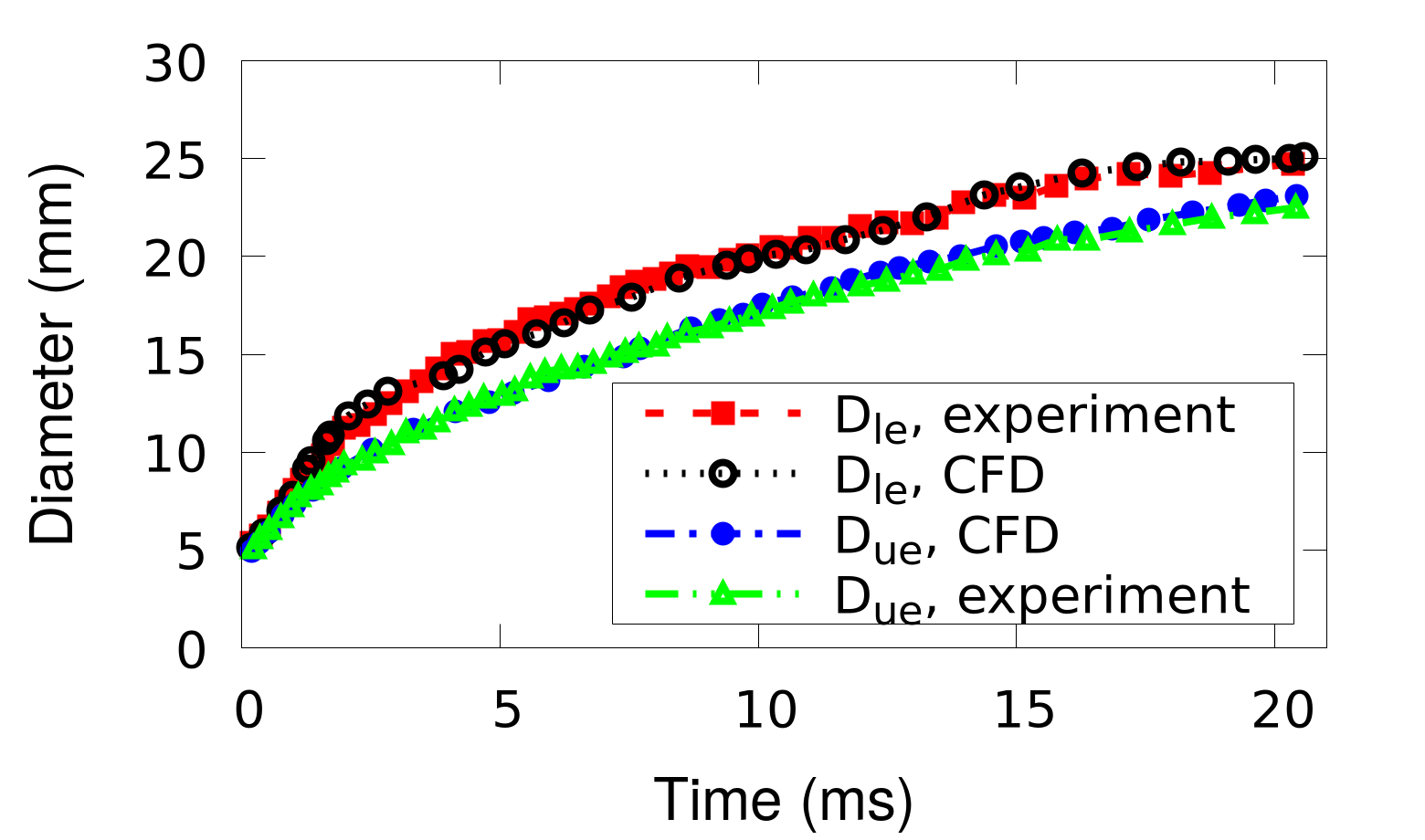}
    \label{fig:sub2}
  \end{subfigure}
\caption{Definition of the $D_{ue}$: crown upper external, $D_{le}$: crown lower external diameters (left) and comparisons of the presented method with the experiments of \cite{cossali2004role} (right).}
\label{fig:drop_impact}
\end{figure}

This test evaluates the capability of the presented method to simulate flows with high Weber (We) and Reynolds (Re) numbers. Available experimental results provide the opportunity to compare the results for the diameter of a splashing drop, providing an appropriate metric for validating the methodology here. This is achieved by calculating the liquid crown diameters at different positions of the impinging drop. 
A droplet of diameter D = 3.82 mm falls on a pre-existing liquid film with thickness 2.3 mm and deforms changing its diameter with time. The Weber number for this case was $We_{D}= 667$ and $Re_{D}$=13676. 
The problem was simulated in 3D using a uniform grid with a mesh size $\Delta$x such that D/$\Delta$x = 44, with approximately 7 million hexahedral cells.
Fig.~\ref{fig:drop_impact}-left shows the definition of the diameters that describe the splash morphology e.g. $D_{ue}$ - crown upper external, $D_{le}$ - crown lower external. These diameters were measured in the experiments of \cite{cossali2004role} and are compared here to the results obtained with the presented solver in Fig.~\ref{fig:drop_impact}-right. 
The simulations showed that both diameters increase with time for the first 20 ms. 
In general, excellent agreement with the experiments was observed with the method used here, demonstrating the capability of the solver to successfully model the evolution of the interface dynamics in problems with high We and Re.

\subsection{Application to liquid jet atomisation}
An application of the methodology is presented for simulating liquid atomisation for high-speed liquid jets in moderate pressure diesel injectors. The test case is taken from the DNS simulations of \cite{chesnel2010simulation} and has a rectangular parallelepiped domain with dimensions 2.5mm $\times$ 2mm $\times$  2mm. The physical parameters of the simulations are shown in Table~\ref{tab:spray}. The liquid jet emerges from the bottom of the domain, from an orifice with diameter D placed in (0,0,0). The liquid is injected at a velocity of $79m/s$.  

The Eulerian-Lagrangian-Spray-Atomisation (ELSA) method \citep{Vallet2001} is employed for modelling the emerging spray. The method solves an equation for the liquid volume fraction and an equation for the surface density denoted $\Sigma$. Various modifications for the formula of $\Sigma$ have been proposed as in \cite{Menard2007,Lebas2009,Lyras2018}. In its general form, the surface density equation includes terms that account for the surface generation or destruction during the atomisation. \cite{Vallet1999} proposed a model which includes the effect of the involved processes in $\Sigma$. The model can be written for different source terms for the contribution of the changes in surface density due to turbulence, aerodynamic break-up and evaporation, and the equation implemented in the code MPflow here is as in \cite{Lyras2018}
\begin{equation} 
\begin{aligned}
& \frac{\partial \bar{\Sigma}}{\partial t} + \frac{\partial \tilde{u_{j}} \bar {\Sigma}}{\partial x_{j}} =                  
   \frac{\partial}{\partial x_{j}}\left(\frac{\nu_{t}}{Sc_{t}}\frac{\partial \bar {\Sigma}}{\partial   x_{j}}\right)+\Psi\left(S_{init} + S_{turb}+S_{vap,den} \right)+ \\
& \left(1-\Psi \right) \left(S_{coll} + S_{2ndBU} + S_{vap,dil}\right) 
\end{aligned}
\label{ref:4.53-equation}
\end{equation}
With $\bar{\Sigma}$ the Reynolds average and $\tilde{u_{j}}$ the mass weighted Favre-average of velocity. The terms on the right-hand side are split for the dense and dilute part of the spray using an indicator function $\Psi$ which is equal to 1 if the liquid mass fraction, $\tilde{Y_{l}}$ is between 1/2 and 1 and becomes zero for cells with a liquid mass fraction less than 0.1. The indicator function can be written as a function of the updated liquid volume fraction, $\alpha$ as,
\begin{equation}
\Psi(\alpha)=H(\alpha-0.1)H(\alpha-0.5)+ (H(\alpha-0.1)-H(\alpha-0.5))(2.5\alpha-0.25)
\label{4.34-equation}
\end{equation}
Similarly to \citep{Vallet1999}, the terms on the right-hand side are written as 
\begin{equation}
S=\frac{\bar{\Sigma}}{\tau_{\Sigma}}\left( 1- \frac{\bar{\Sigma}}{\bar{\Sigma}_{eq}} \right)
\label{4.35-equation}
\end{equation}
where $\bar{\Sigma}_{eq},\tau_{\Sigma}$ are an equilibrium value for the interface and the time-scale of the corresponding process. 

The minimum liquid-gas surface generated in the primary atomisation process $S_{init}$ is proportional to the gradient of liquid mass fraction and is proportional to the inverse of the integration kernel and the characteristic turbulent spatial scales $S_{init} = \alpha(1-\alpha)/l_t$ (\cite{Menard2007}, assuming the first blobs during primary atomisation have the size of $l_t$). If the liquid mass fraction becomes small, and $Y_l(1-Y_l) \leq 0.001$, $S_{init}$ is a function of the gradient of liquid mass fraction obtained from $\alpha=\bar{\rho}\tilde{Y_{l}}/\bar{\rho_{l}}$, and becomes $S_{init}= 2\frac{\mu_{t}}{Sc_{t}} \frac{6 \bar{\rho}}{\rho_{l}\rho_{g}l_{t}}\frac{\partial \tilde{Y_{l}}}{\partial x_{i}} \frac{\partial \tilde{Y_{l}}}{\partial x_{i}}$ \citep{Lebas2009}. The term $S_{turb}$ for the production/destruction of the interface density due to surface stretching due to turbulence in the dense part of the spray is calculated using an equation of the form of Eq.\eqref{4.35-equation} for an equilibrium value, $\Sigma^{*}_{turb}$ which defined from an equilibrium Weber number $We^*$, $We^*=\rho_l \alpha k /\sigma \Sigma^{*}_{turb}$ which is set to 1 for the tests here. The turbulent time-scale $\tau_{t}$ for large-eddy-simulations (LES) the turbulent time-scale is $\tau_{t}=||S_{ij}||^{-1}$, where $S_{ij}$ is the strain rate tensor. Only the primary atomisation terms are considered here. More details for the terms are provided in \cite{Lyras2018}.  

Two meshes with 256 $\times$ 64 $\times$ 64 cells (coarse) and 256 $\times$ 128 $\times$ 128 cells (fine) were used for the tests. An expansion ratio of 3 is used for refining the meshes close to the injection, with a cell size for the smallest cells equal to 10$\mu$m in the axial x-direction, and 5$\mu$m cells in the y and z-directions for the finer mesh. 
The meshes used here, are stretched towards the shear of the jet. The minimum droplet size in the simulations were comparable to the grid size (5$\mu$m for the finest mesh). The size of the droplets depends on the mesh resolution. The DNS studies, with a finer mesh, have shown that this can as small as 2.5$\mu$m. The turbulent length scale ($l_t$) of the DNS simulation was 10$\mu$m with an inflow with a turbulence intensity of 5$\%$. In the present LES, neither of the meshes had enough resolution to accurately capture such small turbulent fluctuations as $\Delta \approx l_t$, (where $\Delta$ is the LES filter) but we are aiming to resolve most of the energy spectrum \citep{pope2001turbulent}. 
For this reason, the ratio of the integral length scale to the grid length scale, $\Delta$ which is taken as the cubic root of the local volume cell, is 10 for the coarse and 20 for the fine mesh \citep{pope2001turbulent}. 
Similarly, we use the same mesh resolution for the simulations here as other LES studies such as in \cite{NavarroMartinez2014} who used the same test case.

\begin{figure}[!tbp]
  \begin{subfigure}[b]{0.62\textwidth}
    \includegraphics[width=\textwidth]{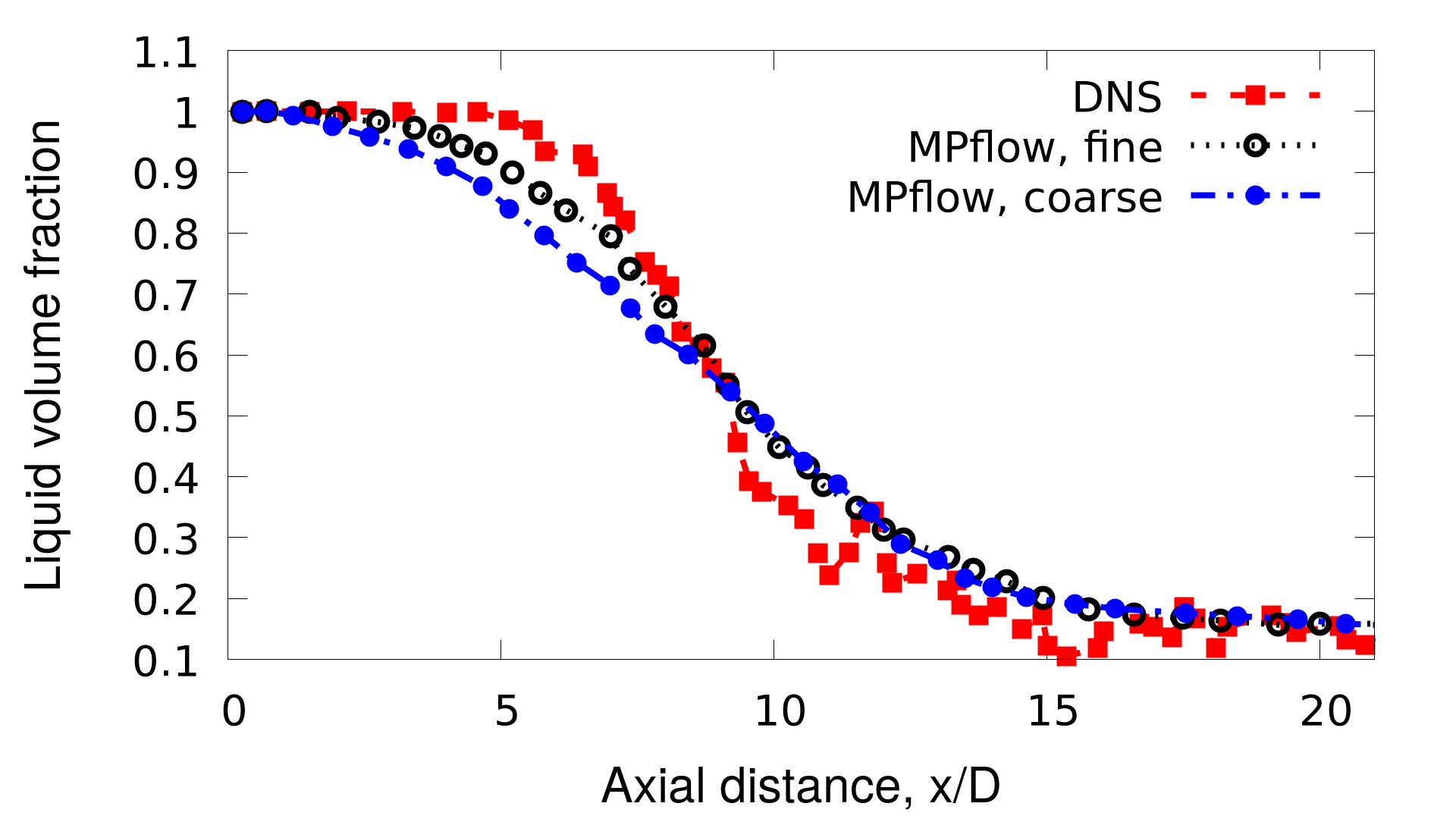}
    \caption{Liquid volume fraction.}
    \label{fig:sub1}
  \end{subfigure}
  \hfill
  \begin{subfigure}[b]{0.62\textwidth}
    \includegraphics[width=\textwidth]{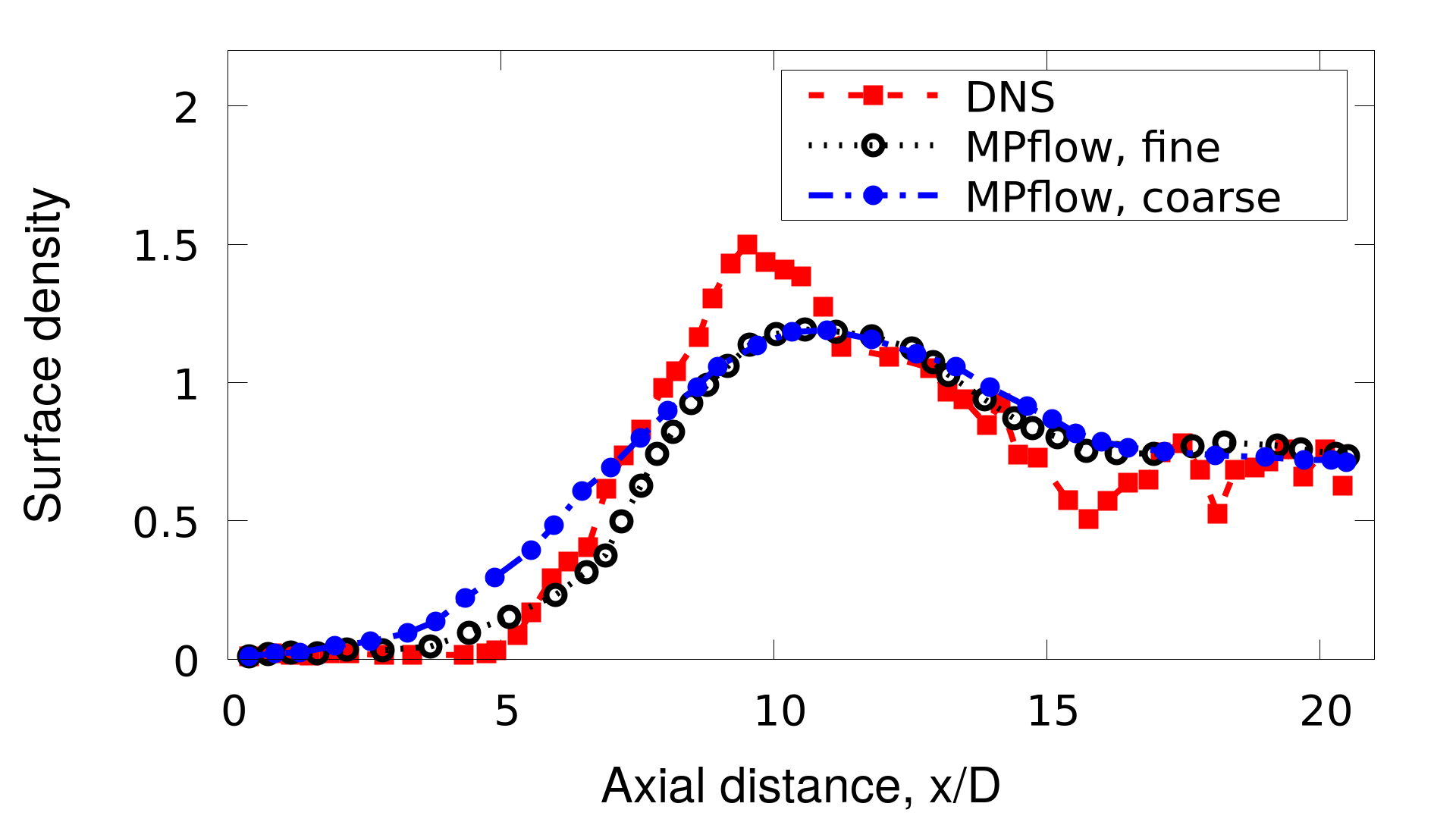}
    \caption{Surface density.}
    \label{fig:sub2}
  \end{subfigure}
\caption{Axial distribution of liquid volume fraction and surface density. Symbols indicate DNS data from surface density ($\Sigma/\Sigma_{0}$).}
\label{fig:alpha_sigma_axial}
\end{figure}

LES simulations were performed using the Smagorinsky model and the sub-grid-scale Reynolds stress, $\tau^{sgs}$ was modelled according to $\tau^{sgs}_{ij}-1/3\tau^{sgs}_{kk}\delta_{ij}=2\mu_t\bar{S}_{ij}$. The sub-grid-scale eddy viscosity was taken to be $\mu_t=C^2_S \rho \Delta^2\parallel\bar{S}\parallel$, where $\parallel \bar{S}\parallel =(\bar{S}_{ij}\bar{S}_{ij})^{1/2}$. For the cases here, the constant $C_S$ was taken equal to 0.1.
The liquid volume fraction at the axial direction is shown in Fig.~\ref{fig:alpha_sigma_axial}a) for both meshes showing similar trends with the DNS results.  
Results in Fig.~\ref{fig:alpha_sigma_axial}b) of the mean surface density at the axial direction are normalised with a theoretical initial value  $\Sigma_0 = 1/\Delta^3$ where $\Delta$ is the LES filter used in the simulations. This is the value that $\Sigma$ will scale with a very fine mesh \citep{NavarroMartinez2014}.
The coarse mesh results demonstrate two regions for the liquid volume fraction, one wherein the prediction for $\alpha$ is underpredicted up to $18\%$, and a second region at $x/D \geq$ 9 where the code overpredicts liquid volume fraction with $5-10\%$ difference. The results for the finer mesh were better for the axial predictions of liquid volume fraction, indicating a better resolution of the liquid penetration and primary atomisation. The local Weber numbers, denoted with $We_D$ are in average greater than the critical Weber number, $We_{h,crit}$. $We_{h,crit}$ is defined as $C_{\Sigma} \sqrt{(1-\alpha)/\alpha}$, where $C_{\Sigma}=2.4$ is a constant proposed \cite{Chesnel2011}. Here, $We_{h,crit} \geq 4.8$ for the dense region for $\alpha \geq 0.2$. 

\begin{figure}[!tbp] 
 \vspace{6pt}
\centering
\begin{subfigure}{.62\textwidth}
  \centering
  \includegraphics[width=1\linewidth]{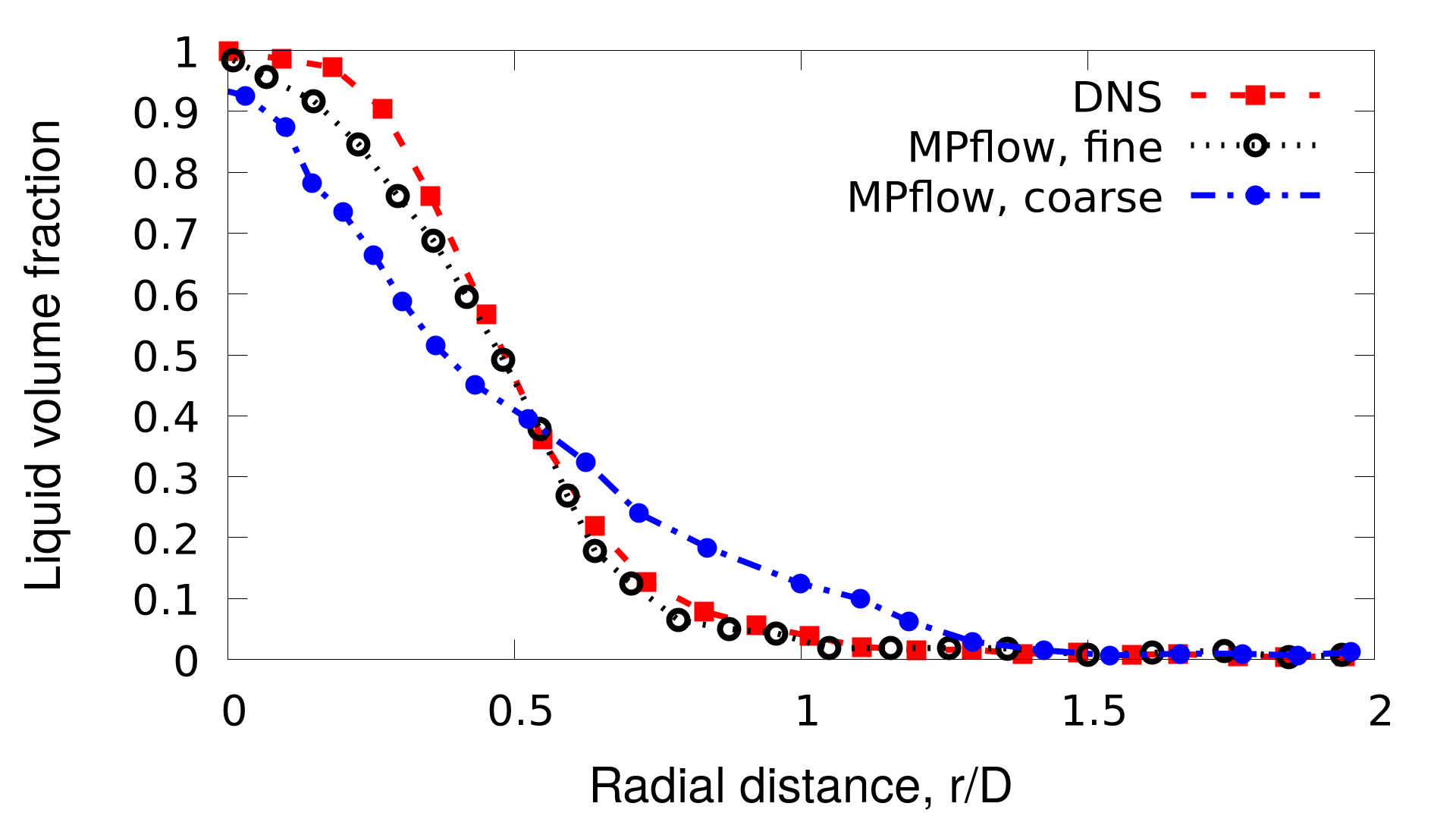}
  \caption{x/D=5}
  \label{fig:sub4}
\end{subfigure}%
\begin{subfigure}{.62\textwidth}
  \centering
  \includegraphics[width=1\linewidth]{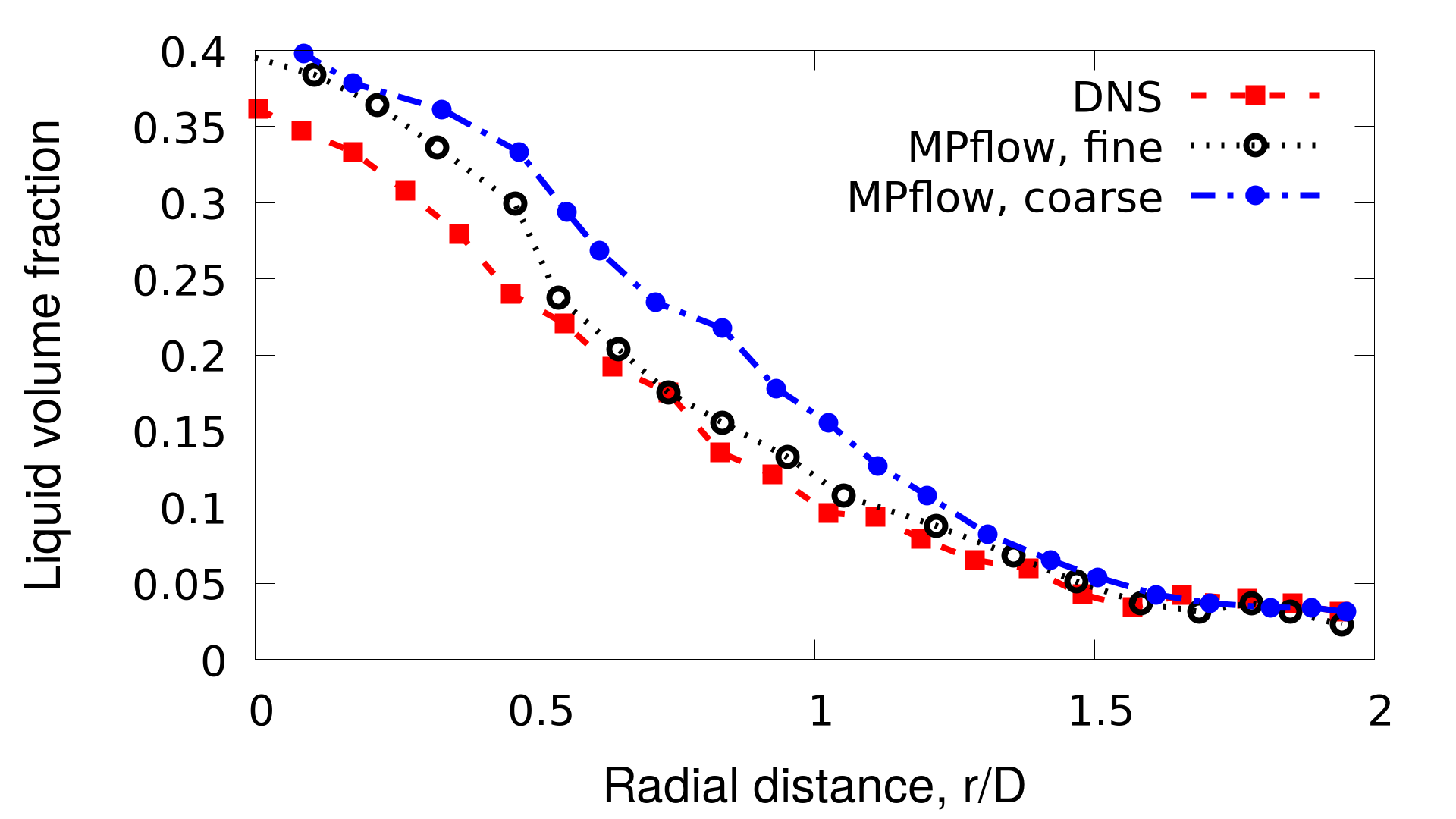}
  \caption{x/D=10}
  \label{fig:sub5}
\end{subfigure}
\begin{subfigure}{.62\textwidth}
  \centering
  \includegraphics[width=1\linewidth]{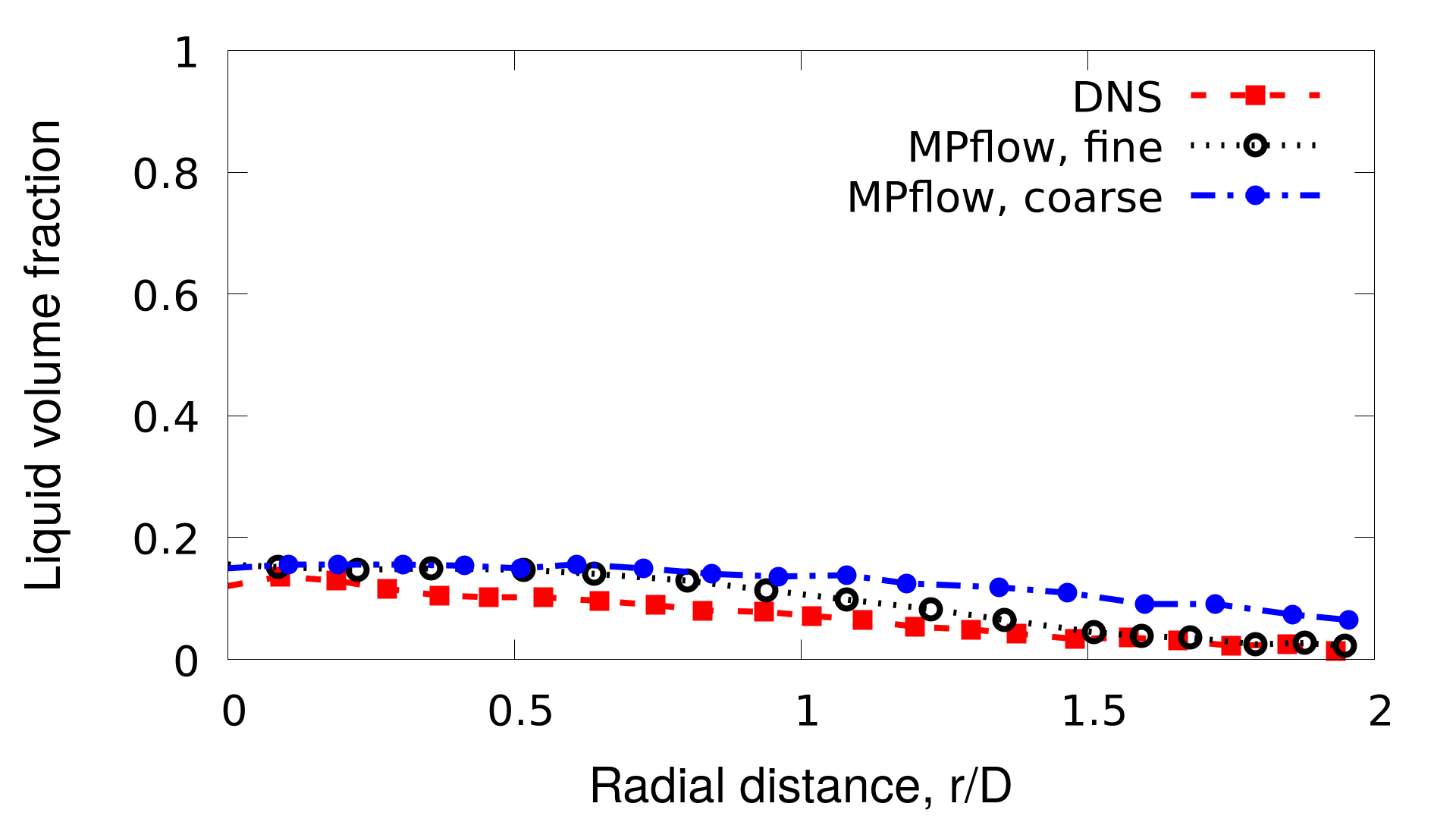}
  \caption{x/D=20}
  \label{fig:sub6}
\end{subfigure}%
\caption{Radial distribution of the liquid volume fraction for the liquid atomisation test for different distances from the centreline.}
\label{fig:alpha_radial}
\end{figure}

\begin{figure}[!tbp]
 \vspace{6pt}
\centering
\begin{subfigure}{.62\textwidth}
  \centering
  \includegraphics[width=1\linewidth]{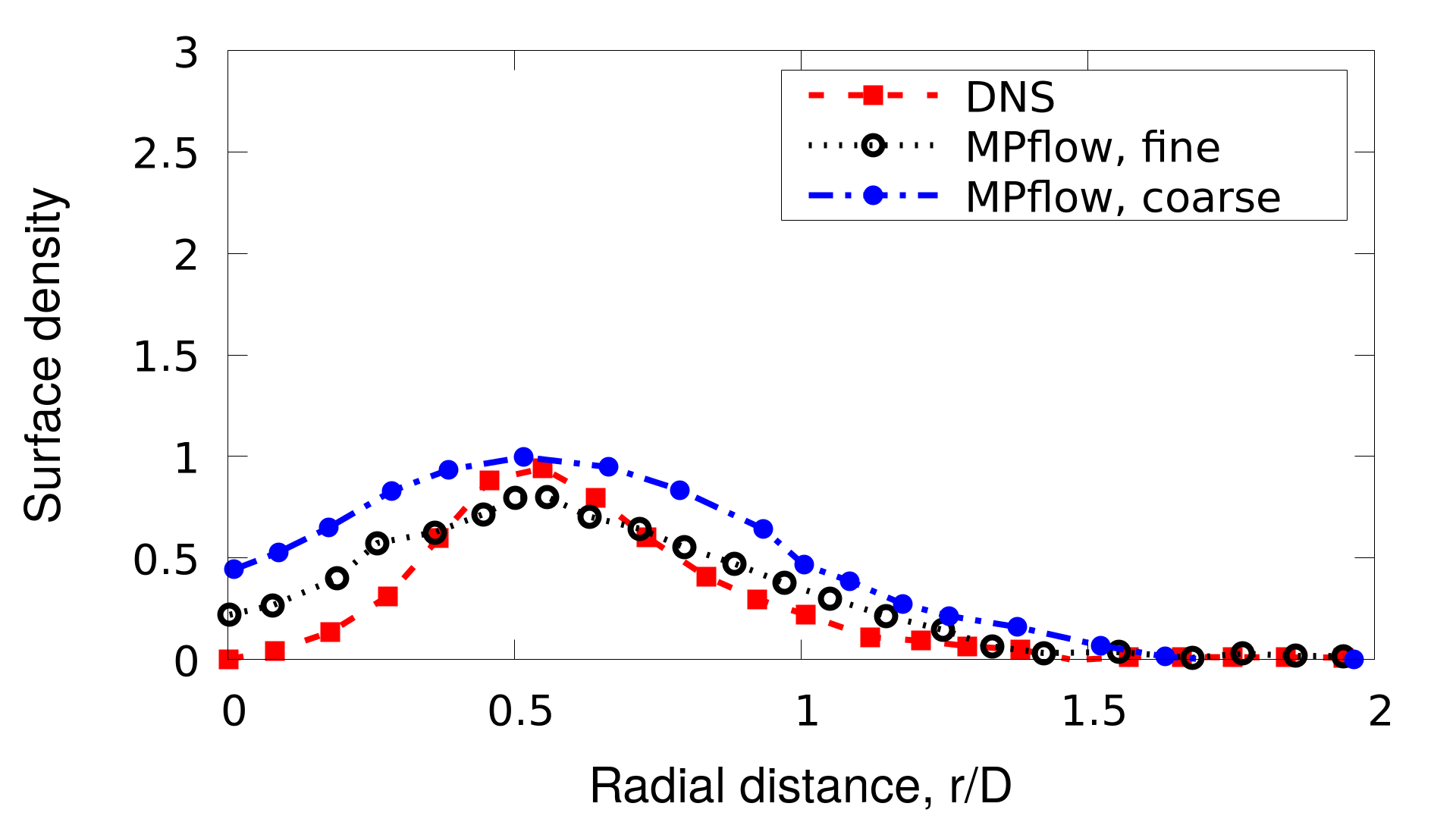}
  \caption{x/D=5}
  \label{fig:sub4}
\end{subfigure}%
\begin{subfigure}{.62\textwidth}
  \centering
  \includegraphics[width=1\linewidth]{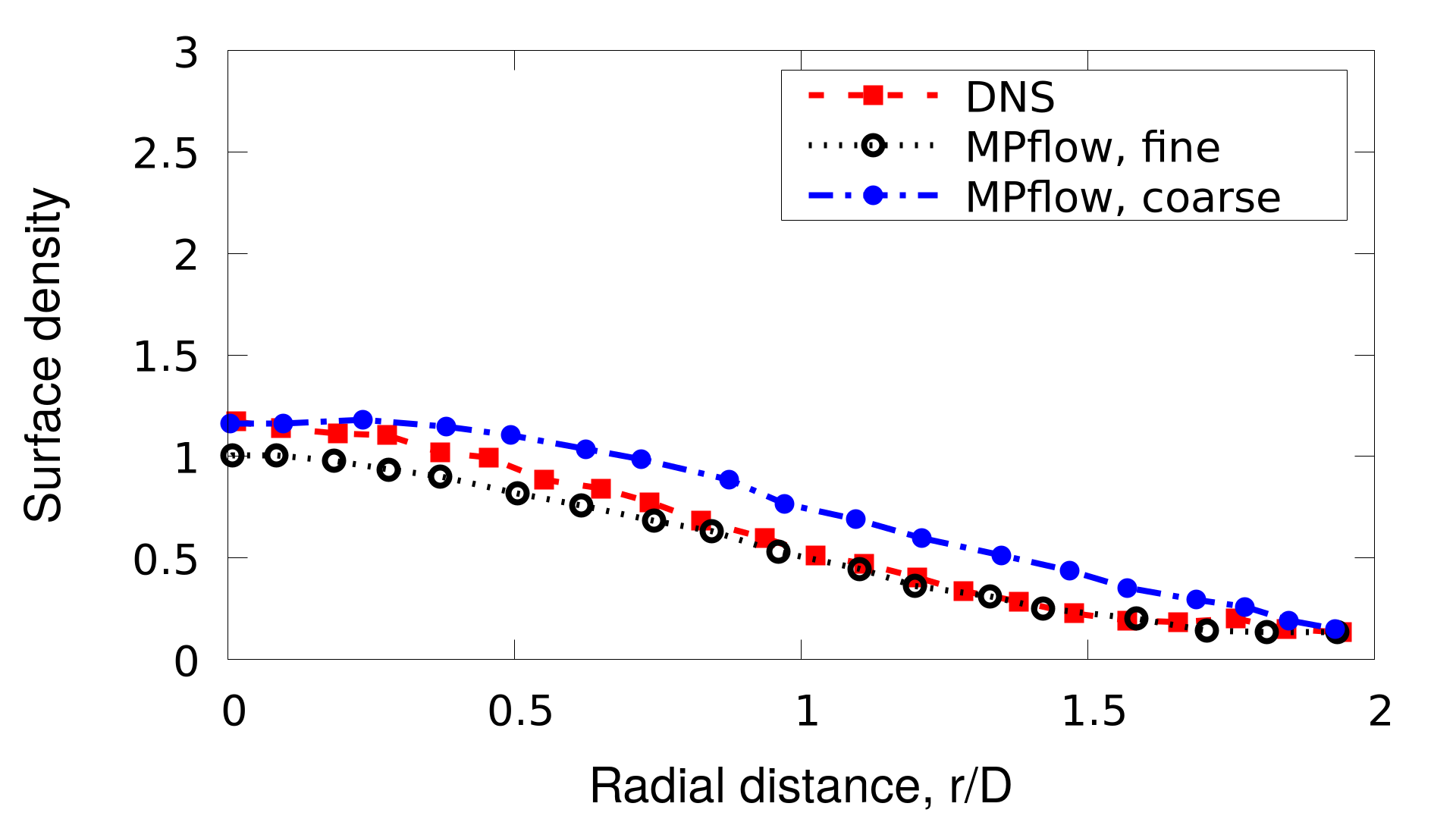}
  \caption{x/D=10}
  \label{fig:sub5}
\end{subfigure}
\begin{subfigure}{.62\textwidth}
  \centering
  \includegraphics[width=1\linewidth]{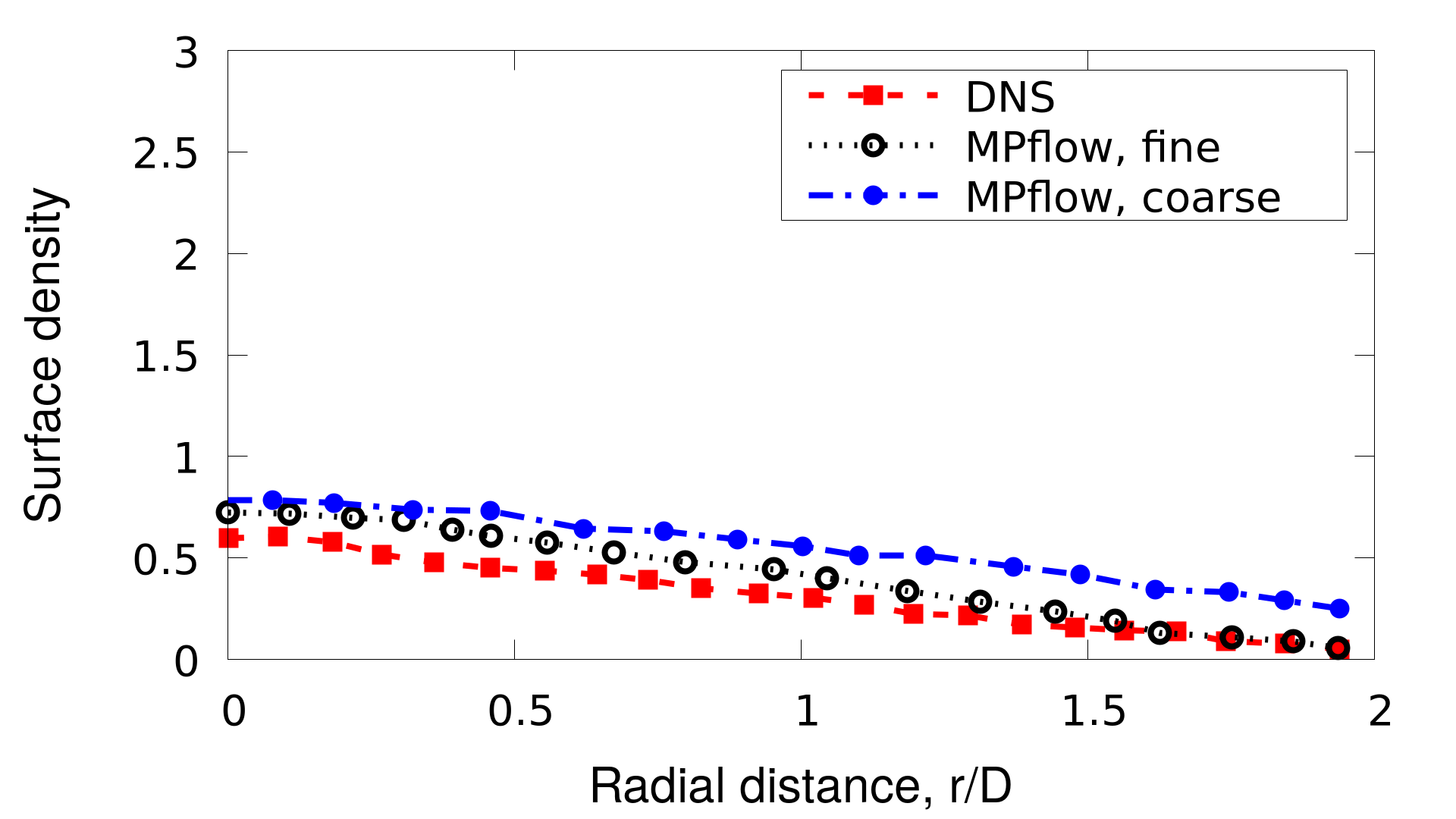}
  \caption{x/D=20}
  \label{fig:sub6}
\end{subfigure}%
\caption{Radial distribution of the surface density for the liquid atomisation test for different distances from the centreline.}
\label{fig:sigma_radial}
\end{figure}

The liquid volume fraction and surface density at the radial direction at x/D=5, x/D=10, x/D=20 are shown in Fig.~\ref{fig:alpha_radial} and Fig.~\ref{fig:sigma_radial} respectively for both meshes.
The highest difference with the DNS results were around 9$\leq$ x/D $\leq$11 for the fine LES which leads to an under-prediction in the surface density at the same region. At the rest of the examined points in both the centreline and radial directions, results obtained with the fine LES here, have an overall reasonable deviation with DNS. The profile of the liquid volume fraction in the entire domain and its trends in the primary atomisation and the jet break-up regions follow the results reported in other studies that have studied the particular case in \cite{Chesnel2011, NavarroMartinez2014}.

Under the tested release conditions, both droplets and ligaments shed from the main liquid core jet which is fragmented into a large parcel of smaller droplets. For the minimum droplet size, we assume that no aerodynamic breakup occurs for Weber numbers smaller than the critical value, which defines the minimum cut-off size of the droplets to be similar to the ones in \cite{chesnel2010simulation}. Although there is a good agreement for the normalised $\Sigma$, the peak value at the vicinity of the position x/D $\approx$10 is lower for both meshes compared to the DNS results, which is attributed to the lower grid resolution. 

\begin{table}[h]
\centering
\caption{Physical properties for the primary atomisation simulations.}
\vspace{6pt}
\begin{tabular}{c|ccc}
\toprule
& Physical and numerical parameters  \\
\midrule 
Liquid density &  696$kg/m^3$ \\
Gas density & 25$kg/m^3$ \\
Liquid viscosity &  1.18e-03$kg/ms$ \\
Gas viscosity &  1.0e-05$kg/ms$ \\
Surface tension & 0.06$N/m$  \\
Diameter D & 100$\mu m$    \\
Weber number (liquid), $We_{l}$ & 7239 \\
Reynolds number (liquid), $Re_{l}$ & 4659 \\
\end{tabular}
\label{tab:spray}
\end{table}

\section{Conclusions}
A coupled level set method with volume of fluid is presented suitable for simulating two-phase flows which can be used for capturing the interface between two fluids.
The methodology is based on an advection-correction step, within the pressure-velocity algorithm, of the level set function which is solved with a high-order scheme and then coupled with volume of fluid. 
The method remains relatively simple to implement and can be coupled with other models for tracking the interface such as the Eulerian-Lagrangian-Spray-Atomisation method here. 
The results showed good accuracy for problems with three-dimensional free-surface phenomena for both structured and unstructured meshes with various mesh resolutions. 
The method is mass-conserving even for coarser meshes and is suitable for long-time simulations. 
Results for primary atomisation of a jet were in close agreement to previously published results for the liquid volume fraction and surface density. 
Although here, a simple mapping step is carried out for mapping the volume fraction to the level set function, a more suitable scheme could be used for obtaining the exact value for the level set which corresponds to the volume of fluid solution that would further improve mass conservation. The added correction steps that introduce additional numerical parameters might have to be further examined in future studies especially when using the method for other applications such as bubbly flows.
   
\section*{Acknowledgments}
The authors would like to thank the Lyras LP for providing the computational resources and working on the co-development of the code MPflow that have contributed to the research results reported within this paper under the project Numerical simulations for complEx fluiD flOw systems desigN (NEDON). 

\bibliographystyle{model1-num-names}
\bibliography{mybibfile}

\begin{thebibliography}{73}
\expandafter\ifx\csname natexlab\endcsname\relax\def\natexlab#1{#1}\fi
\providecommand{\url}[1]{\texttt{#1}}
\providecommand{\href}[2]{#2}
\providecommand{\path}[1]{#1}
\providecommand{\DOIprefix}{doi:}
\providecommand{\ArXivprefix}{arXiv:}
\providecommand{\URLprefix}{URL: }
\providecommand{\Pubmedprefix}{pmid:}
\providecommand{\doi}[1]{\href{http://dx.doi.org/#1}{\path{#1}}}
\providecommand{\Pubmed}[1]{\href{pmid:#1}{\path{#1}}}
\providecommand{\bibinfo}[2]{#2}
\ifx\xfnm\relax \def\xfnm[#1]{\unskip,\space#1}\fi
\bibitem[{Agbaglah et~al.(2011)Agbaglah, Delaux, Fuster, Hoepffner, Josserand,
  Popinet, Ray, Scardovelli and Zaleski}]{Agbaglah2011}
\bibinfo{author}{Agbaglah, G.}, \bibinfo{author}{Delaux, S.},
  \bibinfo{author}{Fuster, D.}, \bibinfo{author}{Hoepffner, J.},
  \bibinfo{author}{Josserand, C.}, \bibinfo{author}{Popinet, S.},
  \bibinfo{author}{Ray, P.}, \bibinfo{author}{Scardovelli, R.},
  \bibinfo{author}{Zaleski, S.}, \bibinfo{year}{2011}.
\newblock \bibinfo{title}{Parallel simulation of multiphase flows using octree
  adaptivity and the volume-of-fluid method}.
\newblock \bibinfo{journal}{Comptes Rendus Mecanique} \bibinfo{volume}{339},
  \bibinfo{pages}{194--207}.
\bibitem[{Albadawi et~al.(2013)Albadawi, Donoghue, Robinson, Murray and
  Delauré}]{Albadawi2013}
\bibinfo{author}{Albadawi, A.}, \bibinfo{author}{Donoghue, D.B.},
  \bibinfo{author}{Robinson, A.J.}, \bibinfo{author}{Murray, D.B.},
  \bibinfo{author}{Delauré, Y.M.C.}, \bibinfo{year}{2013}.
\newblock \bibinfo{title}{Influence of surface tension implementation in volume
  of fluid and coupled volume of fluid with level set methods for bubble growth
  and detachment}.
\newblock \bibinfo{journal}{International Journal of Multiphase Flow}
  \bibinfo{volume}{53}, \bibinfo{pages}{11 -- 28}.
\newblock
  \DOIprefix\doi{https://doi.org/10.1016/j.ijmultiphaseflow.2013.01.005}.
\bibitem[{Arienti and Sussman(2014)}]{Arienti2014}
\bibinfo{author}{Arienti, M.}, \bibinfo{author}{Sussman, M.},
  \bibinfo{year}{2014}.
\newblock \bibinfo{title}{An embedded level set method for sharp-interface
  multiphase simulations of diesel injectors}.
\newblock \bibinfo{journal}{International Journal of Multiphase Flow}
  \bibinfo{volume}{59}, \bibinfo{pages}{1 -- 14}.
\newblock
  \DOIprefix\doi{https://doi.org/10.1016/j.ijmultiphaseflow.2013.10.005}.
\bibitem[{Aulisa et~al.(2003)Aulisa, Manservisi, Scardovelli and
  Zaleski}]{Aulisa2003}
\bibinfo{author}{Aulisa, E.}, \bibinfo{author}{Manservisi, S.},
  \bibinfo{author}{Scardovelli, R.}, \bibinfo{author}{Zaleski, S.},
  \bibinfo{year}{2003}.
\newblock \bibinfo{title}{A geometrical area-preserving volume-of-fluid
  advection method}.
\newblock \bibinfo{journal}{Journal of Computational Physics}
  \bibinfo{volume}{192}, \bibinfo{pages}{355--364}.
\bibitem[{Balc{\'a}zar et~al.(2014)Balc{\'a}zar, Jofre, Lehmkuhl, Castro and
  Rigola}]{Balcazar2014}
\bibinfo{author}{Balc{\'a}zar, N.}, \bibinfo{author}{Jofre, L.},
  \bibinfo{author}{Lehmkuhl, O.}, \bibinfo{author}{Castro, J.},
  \bibinfo{author}{Rigola, J.}, \bibinfo{year}{2014}.
\newblock \bibinfo{title}{A finite-volume/level-set method for simulating
  two-phase flows on unstructured grids}.
\newblock \bibinfo{journal}{International Journal of Multiphase Flow}
  \bibinfo{volume}{64}, \bibinfo{pages}{55 -- 72}.
\newblock
  \DOIprefix\doi{https://doi.org/10.1016/j.ijmultiphaseflow.2014.04.008}.
\bibitem[{Balc{\'a}zar et~al.(2015)Balc{\'a}zar, Lehmkuhl, Rigola and
  Oliva}]{Balcazar2015}
\bibinfo{author}{Balc{\'a}zar, N.}, \bibinfo{author}{Lehmkuhl, O.},
  \bibinfo{author}{Rigola, J.}, \bibinfo{author}{Oliva, A.},
  \bibinfo{year}{2015}.
\newblock \bibinfo{title}{A multiple marker level-set method for simulation of
  deformable fluid particles}.
\newblock \bibinfo{journal}{International Journal of Multiphase Flow}
  \bibinfo{volume}{74}, \bibinfo{pages}{125 -- 142}.
\newblock
  \DOIprefix\doi{https://doi.org/10.1016/j.ijmultiphaseflow.2015.04.009}.
\bibitem[{Bourlioux(1995)}]{Bourlioux1995}
\bibinfo{author}{Bourlioux, A.}, \bibinfo{year}{1995}.
\newblock \bibinfo{title}{A coupled level set and volume of fluid algorithm for
  tracking material interfaces}.
\newblock \bibinfo{journal}{Proceedings of the 6th International Symposium On
  Computational Fluid Dynamics} \bibinfo{volume}{6}, \bibinfo{pages}{15--22}.
\bibitem[{Brackbill et~al.(1992)Brackbill, Kothe and Zemach}]{Brackbill1992}
\bibinfo{author}{Brackbill, J.U.}, \bibinfo{author}{Kothe, D.B.},
  \bibinfo{author}{Zemach, C.}, \bibinfo{year}{1992}.
\newblock \bibinfo{title}{A continuum method for modeling surface tension}.
\newblock \bibinfo{journal}{Journal of computational physics}
  \bibinfo{volume}{100}, \bibinfo{pages}{335--354}.
\bibitem[{Cerquaglia et~al.(2017)Cerquaglia, Deli{\'e}ge, Boman, Terrapon and
  Ponthot}]{Cerquaglia2017}
\bibinfo{author}{Cerquaglia, M.L.}, \bibinfo{author}{Deli{\'e}ge, G.},
  \bibinfo{author}{Boman, R.}, \bibinfo{author}{Terrapon, V.},
  \bibinfo{author}{Ponthot, J.P.}, \bibinfo{year}{2017}.
\newblock \bibinfo{title}{Free-slip boundary conditions for simulating
  free-surface incompressible flows through the particle finite element
  method}.
\newblock \bibinfo{journal}{International Journal for Numerical Methods in
  Engineering} \bibinfo{volume}{110}, \bibinfo{pages}{921--946}.
\bibitem[{Chang et~al.(1996)Chang, Hou, Merriman and Osher}]{Chang1996}
\bibinfo{author}{Chang, Y.C.}, \bibinfo{author}{Hou, T.},
  \bibinfo{author}{Merriman, B.}, \bibinfo{author}{Osher, S.},
  \bibinfo{year}{1996}.
\newblock \bibinfo{title}{A level set formulation of eulerian interface
  capturing methods for incompressible fluid flows}.
\newblock \bibinfo{journal}{Journal of computational Physics}
  \bibinfo{volume}{124}, \bibinfo{pages}{449--464}.
\bibitem[{Chesnel(2010)}]{chesnel2010simulation}
\bibinfo{author}{Chesnel, J.}, \bibinfo{year}{2010}.
\newblock \bibinfo{title}{Simulation aux Grandes {\'E}chelles de l'Atomisation,
  Application {\`a} l'Injection Automobile.}
\newblock Ph.D. thesis. Universit{\'e} de Rouen.
\bibitem[{Chesnel et~al.(2011)Chesnel, Reveillon, Menard and
  Demoulin}]{Chesnel2011}
\bibinfo{author}{Chesnel, J.}, \bibinfo{author}{Reveillon, J.},
  \bibinfo{author}{Menard, T.}, \bibinfo{author}{Demoulin, F.X.},
  \bibinfo{year}{2011}.
\newblock \bibinfo{title}{Large eddy simulation of liquid jet atomization}.
\newblock \bibinfo{journal}{Atomization and Sprays} \bibinfo{volume}{21}.
\bibitem[{Cossali et~al.(2004)Cossali, Marengo, Coghe and
  Zhdanov}]{cossali2004role}
\bibinfo{author}{Cossali, G.}, \bibinfo{author}{Marengo, M.},
  \bibinfo{author}{Coghe, A.}, \bibinfo{author}{Zhdanov, S.},
  \bibinfo{year}{2004}.
\newblock \bibinfo{title}{The role of time in single drop splash on thin film}.
\newblock \bibinfo{journal}{Experiments in Fluids} \bibinfo{volume}{36},
  \bibinfo{pages}{888--900}.
\bibitem[{Deshpande et~al.(2012)Deshpande, Anumolu and
  Trujillo}]{Deshpande2012}
\bibinfo{author}{Deshpande, S.S.}, \bibinfo{author}{Anumolu, L.},
  \bibinfo{author}{Trujillo, M.F.}, \bibinfo{year}{2012}.
\newblock \bibinfo{title}{Evaluating the performance of the two-phase flow
  solver interfoam}.
\newblock \bibinfo{journal}{Computational Science \& Discovery}
  \bibinfo{volume}{5}, \bibinfo{pages}{014016}.
\bibitem[{Dianat et~al.(2017)Dianat, Skarysz and Garmory}]{Dianat2017}
\bibinfo{author}{Dianat, M.}, \bibinfo{author}{Skarysz, M.},
  \bibinfo{author}{Garmory, A.}, \bibinfo{year}{2017}.
\newblock \bibinfo{title}{A coupled level set and volume of fluid method for
  automotive exterior water management applications}.
\newblock \bibinfo{journal}{International Journal of Multiphase Flow}
  \bibinfo{volume}{91}, \bibinfo{pages}{19 -- 38}.
\newblock
  \DOIprefix\doi{https://doi.org/10.1016/j.ijmultiphaseflow.2017.01.008}.
\bibitem[{Duret et~al.(2013)Duret, Reveillon, Menard and Demoulin}]{Duret2013}
\bibinfo{author}{Duret, B.}, \bibinfo{author}{Reveillon, J.},
  \bibinfo{author}{Menard, T.}, \bibinfo{author}{Demoulin, F.},
  \bibinfo{year}{2013}.
\newblock \bibinfo{title}{Improving primary atomization modeling through dns of
  two-phase flows}.
\newblock \bibinfo{journal}{International Journal of Multiphase Flow}
  \bibinfo{volume}{55}, \bibinfo{pages}{130--137}.
\bibitem[{Elias and Coutinho(2007)}]{Elias2007}
\bibinfo{author}{Elias, R.N.}, \bibinfo{author}{Coutinho, A.L.},
  \bibinfo{year}{2007}.
\newblock \bibinfo{title}{Stabilized edge-based finite element simulation of
  free-surface flows}.
\newblock \bibinfo{journal}{International Journal for Numerical Methods in
  Fluids} \bibinfo{volume}{54}, \bibinfo{pages}{965--993}.
\bibitem[{Enright et~al.(2002)Enright, Fedkiw, Ferziger and
  Mitchell}]{Enright2002}
\bibinfo{author}{Enright, D.}, \bibinfo{author}{Fedkiw, R.},
  \bibinfo{author}{Ferziger, J.}, \bibinfo{author}{Mitchell, I.},
  \bibinfo{year}{2002}.
\newblock \bibinfo{title}{A hybrid particle level set method for improved
  interface capturing}.
\newblock \bibinfo{journal}{Journal of Computational physics}
  \bibinfo{volume}{183}, \bibinfo{pages}{83--116}.
\bibitem[{Ferrari et~al.(2017)Ferrari, Magnini and Thome}]{Ferrari2017}
\bibinfo{author}{Ferrari, A.}, \bibinfo{author}{Magnini, M.},
  \bibinfo{author}{Thome, J.R.}, \bibinfo{year}{2017}.
\newblock \bibinfo{title}{A flexible coupled level set and volume of fluid
  (flexclv) method to simulate microscale two-phase flow in non-uniform and
  unstructured meshes}.
\newblock \bibinfo{journal}{International Journal of Multiphase Flow}
  \bibinfo{volume}{91}, \bibinfo{pages}{276 -- 295}.
\newblock
  \DOIprefix\doi{https://doi.org/10.1016/j.ijmultiphaseflow.2017.01.017}.
\bibitem[{Fuster et~al.(2009)Fuster, Bagué, Boeck, {Le Moyne}, Leboissetier,
  Popinet, Ray, Scardovelli and Zaleski}]{Fuster2009}
\bibinfo{author}{Fuster, D.}, \bibinfo{author}{Bagué, A.},
  \bibinfo{author}{Boeck, T.}, \bibinfo{author}{{Le Moyne}, L.},
  \bibinfo{author}{Leboissetier, A.}, \bibinfo{author}{Popinet, S.},
  \bibinfo{author}{Ray, P.}, \bibinfo{author}{Scardovelli, R.},
  \bibinfo{author}{Zaleski, S.}, \bibinfo{year}{2009}.
\newblock \bibinfo{title}{Simulation of primary atomization with an octree
  adaptive mesh refinement and vof method}.
\newblock \bibinfo{journal}{International Journal of Multiphase Flow}
  \bibinfo{volume}{35}, \bibinfo{pages}{550 -- 565}.
\newblock
  \DOIprefix\doi{https://doi.org/10.1016/j.ijmultiphaseflow.2009.02.014}.
\bibitem[{Gottlieb and Shu(1998)}]{gottlieb1998total}
\bibinfo{author}{Gottlieb, S.}, \bibinfo{author}{Shu, C.W.},
  \bibinfo{year}{1998}.
\newblock \bibinfo{title}{Total variation diminishing runge-kutta schemes}.
\newblock \bibinfo{journal}{Mathematics of computation} \bibinfo{volume}{67},
  \bibinfo{pages}{73--85}.
\bibitem[{Gueyffier et~al.(1999)Gueyffier, Li, Nadim, Scardovelli and
  Zaleski}]{Gueyffier1999}
\bibinfo{author}{Gueyffier, D.}, \bibinfo{author}{Li, J.},
  \bibinfo{author}{Nadim, A.}, \bibinfo{author}{Scardovelli, R.},
  \bibinfo{author}{Zaleski, S.}, \bibinfo{year}{1999}.
\newblock \bibinfo{title}{Volume-of-fluid interface tracking with smoothed
  surface stress methods for three-dimensional flows}.
\newblock \bibinfo{journal}{Journal of Computational physics}
  \bibinfo{volume}{152}, \bibinfo{pages}{423--456}.
\bibitem[{Haghshenas et~al.(2017)Haghshenas, Wilson and Kumar}]{Haghshenas2017}
\bibinfo{author}{Haghshenas, M.}, \bibinfo{author}{Wilson, J.A.},
  \bibinfo{author}{Kumar, R.}, \bibinfo{year}{2017}.
\newblock \bibinfo{title}{Algebraic coupled level set-volume of fluid method
  for surface tension dominant two-phase flows}.
\newblock \bibinfo{journal}{International Journal of Multiphase Flow}
  \bibinfo{volume}{90}, \bibinfo{pages}{13 -- 28}.
\newblock
  \DOIprefix\doi{https://doi.org/10.1016/j.ijmultiphaseflow.2016.12.002}.
\bibitem[{Hartmann et~al.(2008)Hartmann, Meinke and
  Schr{\"o}der}]{Hartmann2008}
\bibinfo{author}{Hartmann, D.}, \bibinfo{author}{Meinke, M.},
  \bibinfo{author}{Schr{\"o}der, W.}, \bibinfo{year}{2008}.
\newblock \bibinfo{title}{Differential equation based constrained
  reinitialization for level set methods}.
\newblock \bibinfo{journal}{Journal of Computational Physics}
  \bibinfo{volume}{227}, \bibinfo{pages}{6821 -- 6845}.
\newblock \DOIprefix\doi{https://doi.org/10.1016/j.jcp.2008.03.040}.
\bibitem[{Hartmann et~al.(2010)Hartmann, Meinke and
  Schr{\"o}der}]{Hartmann2010}
\bibinfo{author}{Hartmann, D.}, \bibinfo{author}{Meinke, M.},
  \bibinfo{author}{Schr{\"o}der, W.}, \bibinfo{year}{2010}.
\newblock \bibinfo{title}{The constrained reinitialization equation for level
  set methods}.
\newblock \bibinfo{journal}{Journal of Computational Physics}
  \bibinfo{volume}{229}, \bibinfo{pages}{1514--1535}.
\bibitem[{Hern{\'a}ndez et~al.(2008)Hern{\'a}ndez, L{\'o}pez, G{\'o}mez, Zanzi
  and Faura}]{Hernandez2008}
\bibinfo{author}{Hern{\'a}ndez, J.}, \bibinfo{author}{L{\'o}pez, J.},
  \bibinfo{author}{G{\'o}mez, P.}, \bibinfo{author}{Zanzi, C.},
  \bibinfo{author}{Faura, F.}, \bibinfo{year}{2008}.
\newblock \bibinfo{title}{A new volume of fluid method in three
  dimensions—part i: Multidimensional advection method with face-matched flux
  polyhedra}.
\newblock \bibinfo{journal}{International Journal for Numerical Methods in
  Fluids} \bibinfo{volume}{58}, \bibinfo{pages}{897--921}.
\bibitem[{Hirt and Nichols(1981)}]{Hirt1981}
\bibinfo{author}{Hirt, C.W.}, \bibinfo{author}{Nichols, B.D.},
  \bibinfo{year}{1981}.
\newblock \bibinfo{title}{Volume of fluid (vof) method for the dynamics of free
  boundaries}.
\newblock \bibinfo{journal}{Journal of Computational Physics}
  \bibinfo{volume}{39}, \bibinfo{pages}{201 -- 225}.
\newblock \DOIprefix\doi{https://doi.org/10.1016/0021-9991(81)90145-5}.
\bibitem[{Issa et~al.(1986)Issa, Gosman and Watkins}]{Issa1986}
\bibinfo{author}{Issa, R.}, \bibinfo{author}{Gosman, A.D.},
  \bibinfo{author}{Watkins, A.P.}, \bibinfo{year}{1986}.
\newblock \bibinfo{title}{The computation of compressible and incompressible
  recirculating flows by a non-iterative implicit scheme}.
\newblock \bibinfo{journal}{J. Comput. Phys.} \bibinfo{volume}{62},
  \bibinfo{pages}{66--82}.
\newblock \DOIprefix\doi{10.1016/0021-9991(86)90100-2}.
\bibitem[{Jemison et~al.(2013)Jemison, Loch, Sussman, Shashkov, Arienti, Ohta
  and Wang}]{Jemison2013}
\bibinfo{author}{Jemison, M.}, \bibinfo{author}{Loch, E.},
  \bibinfo{author}{Sussman, M.}, \bibinfo{author}{Shashkov, M.},
  \bibinfo{author}{Arienti, M.}, \bibinfo{author}{Ohta, M.},
  \bibinfo{author}{Wang, Y.}, \bibinfo{year}{2013}.
\newblock \bibinfo{title}{A coupled level set-moment of fluid method for
  incompressible two-phase flows}.
\newblock \bibinfo{journal}{Journal of Scientific Computing}
  \bibinfo{volume}{54}, \bibinfo{pages}{454--491}.
\bibitem[{Kees et~al.(2011)Kees, Akkerman, Farthing and Bazilevs}]{Kees2011}
\bibinfo{author}{Kees, C.E.}, \bibinfo{author}{Akkerman, I.},
  \bibinfo{author}{Farthing, M.W.}, \bibinfo{author}{Bazilevs, Y.},
  \bibinfo{year}{2011}.
\newblock \bibinfo{title}{A conservative level set method suitable for
  variable-order approximations and unstructured meshes}.
\newblock \bibinfo{journal}{J. Comput. Phys.} \bibinfo{volume}{230},
  \bibinfo{pages}{4536--4558}.
\newblock \DOIprefix\doi{10.1016/j.jcp.2011.02.030}.
\bibitem[{Koshizuka and Oka(1996)}]{Koshizuka1996}
\bibinfo{author}{Koshizuka, S.}, \bibinfo{author}{Oka, Y.},
  \bibinfo{year}{1996}.
\newblock \bibinfo{title}{Moving-particle semi-implicit method for
  fragmentation of incompressible fluid}.
\newblock \bibinfo{journal}{Nuclear Science and Engineering}
  \bibinfo{volume}{123}, \bibinfo{pages}{421--434}.
\newblock \DOIprefix\doi{10.13182/NSE96-A24205},
  \href{http://arxiv.org/abs/https://doi.org/10.13182/NSE96-A24205}{\tt
  arXiv:https://doi.org/10.13182/NSE96-A24205}.
\bibitem[{Lafaurie et~al.(1994)Lafaurie, Nardone, Scardovelli, Zaleski and
  Zanetti}]{Lafaurie1994}
\bibinfo{author}{Lafaurie, B.}, \bibinfo{author}{Nardone, C.},
  \bibinfo{author}{Scardovelli, R.}, \bibinfo{author}{Zaleski, S.},
  \bibinfo{author}{Zanetti, G.}, \bibinfo{year}{1994}.
\newblock \bibinfo{title}{Modelling merging and fragmentation in multiphase
  flows with surfer}.
\newblock \bibinfo{journal}{Journal of Computational Physics}
  \bibinfo{volume}{113}, \bibinfo{pages}{134--147}.
\bibitem[{Lakdawala et~al.(2014)Lakdawala, Gada and Sharma}]{Lakdawala2014}
\bibinfo{author}{Lakdawala, A.M.}, \bibinfo{author}{Gada, V.H.},
  \bibinfo{author}{Sharma, A.}, \bibinfo{year}{2014}.
\newblock \bibinfo{title}{A dual grid level set method based study of
  interface-dynamics for a liquid jet injected upwards into another liquid}.
\newblock \bibinfo{journal}{International Journal of Multiphase Flow}
  \bibinfo{volume}{59}, \bibinfo{pages}{206 -- 220}.
\newblock
  \DOIprefix\doi{https://doi.org/10.1016/j.ijmultiphaseflow.2013.11.009}.
\bibitem[{Lebas et~al.(2009)Lebas, Menard, Beau, Berlemont and
  Demoulin}]{Lebas2009}
\bibinfo{author}{Lebas, R.}, \bibinfo{author}{Menard, T.},
  \bibinfo{author}{Beau, P.A.}, \bibinfo{author}{Berlemont, A.},
  \bibinfo{author}{Demoulin, F.X.}, \bibinfo{year}{2009}.
\newblock \bibinfo{title}{Numerical simulation of primary break-up and
  atomization: Dns and modelling study}.
\newblock \bibinfo{journal}{International Journal of Multiphase Flow}
  \bibinfo{volume}{35}, \bibinfo{pages}{247--260}.
\newblock
  \DOIprefix\doi{https://doi.org/10.1016/j.ijmultiphaseflow.2008.11.005}.
\bibitem[{LeVeque(1996)}]{LeVeque1996}
\bibinfo{author}{LeVeque, R.J.}, \bibinfo{year}{1996}.
\newblock \bibinfo{title}{High-resolution conservative algorithms for advection
  in incompressible flow}.
\newblock \bibinfo{journal}{SIAM J. Numer. Anal.} \bibinfo{volume}{33},
  \bibinfo{pages}{627--665}.
\newblock \DOIprefix\doi{10.1137/0733033}.
\bibitem[{Li et~al.(2002)Li, Yu and Chen}]{li2002third}
\bibinfo{author}{Li, X.G.}, \bibinfo{author}{Yu, X.J.}, \bibinfo{author}{Chen,
  G.N.}, \bibinfo{year}{2002}.
\newblock \bibinfo{title}{The third-order relaxation schemes for hyperbolic
  conservation laws}.
\newblock \bibinfo{journal}{Journal of computational and applied mathematics}
  \bibinfo{volume}{138}, \bibinfo{pages}{93--108}.
\bibitem[{Liovic et~al.(2006)Liovic, Rudman, Liow, Lakehal and
  Kothe}]{liovic2006}
\bibinfo{author}{Liovic, P.}, \bibinfo{author}{Rudman, M.},
  \bibinfo{author}{Liow, J.L.}, \bibinfo{author}{Lakehal, D.},
  \bibinfo{author}{Kothe, D.}, \bibinfo{year}{2006}.
\newblock \bibinfo{title}{A 3d unsplit-advection volume tracking algorithm with
  planarity-preserving interface reconstruction}.
\newblock \bibinfo{journal}{Computers \& fluids} \bibinfo{volume}{35},
  \bibinfo{pages}{1011--1032}.
\bibitem[{Liu et~al.(1994)Liu, Osher and Chan}]{Liu1994}
\bibinfo{author}{Liu, X.D.}, \bibinfo{author}{Osher, S.},
  \bibinfo{author}{Chan, T.}, \bibinfo{year}{1994}.
\newblock \bibinfo{title}{Weighted essentially non-oscillatory schemes}.
\newblock \bibinfo{journal}{Journal of Computational Physics}
  \bibinfo{volume}{115}, \bibinfo{pages}{200--212}.
\newblock \DOIprefix\doi{https://doi.org/10.1006/jcph.1994.1187}.
\bibitem[{Lopez et~al.(2005)Lopez, Hernandez, Gomez and Faura}]{Lopez2005}
\bibinfo{author}{Lopez, J.}, \bibinfo{author}{Hernandez, J.},
  \bibinfo{author}{Gomez, P.}, \bibinfo{author}{Faura, F.},
  \bibinfo{year}{2005}.
\newblock \bibinfo{title}{An improved plic-vof method for tracking thin fluid
  structures in incompressible two-phase flows}.
\newblock \bibinfo{journal}{Journal of Computational Physics}
  \bibinfo{volume}{208}, \bibinfo{pages}{51--74}.
\bibitem[{Lyras et~al.(2018)Lyras, Dembele, Schmidt and Wen}]{Lyras2018}
\bibinfo{author}{Lyras, K.}, \bibinfo{author}{Dembele, S.},
  \bibinfo{author}{Schmidt, D.P.}, \bibinfo{author}{Wen, J.X.},
  \bibinfo{year}{2018}.
\newblock \bibinfo{title}{Numerical simulation of subcooled and superheated
  jets under thermodynamic non-equilibrium}.
\newblock \bibinfo{journal}{International Journal of Multiphase Flow}
  \bibinfo{volume}{102}, \bibinfo{pages}{16--28}.
\bibitem[{Lyras et~al.(2020)Lyras, Hanson, Fairweather and Heggs}]{Lyras2020}
\bibinfo{author}{Lyras, K.G.}, \bibinfo{author}{Hanson, B.},
  \bibinfo{author}{Fairweather, M.}, \bibinfo{author}{Heggs, P.J.},
  \bibinfo{year}{2020}.
\newblock \bibinfo{title}{A coupled level set and volume of fluid method with a
  re-initialisation step suitable for unstructured meshes}.
\newblock \bibinfo{journal}{Journal of Computational Physics} ,
  \bibinfo{pages}{109224}.
\bibitem[{Martin and Moyce(1952)}]{Martin1952}
\bibinfo{author}{Martin, J.}, \bibinfo{author}{Moyce, W.},
  \bibinfo{year}{1952}.
\newblock \bibinfo{title}{An experimental study of the collapse of fluid
  columns on a rigid horizontal plane, in a medium of lower, but comparable,
  density. 5.}
\newblock \bibinfo{journal}{Philosophical Transactions of the Royal Society of
  London Series A-Mathematical and Physical Sciences} \bibinfo{volume}{244},
  \bibinfo{pages}{325--334}.
\bibitem[{M{\'e}nard et~al.(2007)M{\'e}nard, Tanguy and Berlemont}]{Menard2007}
\bibinfo{author}{M{\'e}nard, T.}, \bibinfo{author}{Tanguy, S.},
  \bibinfo{author}{Berlemont, A.}, \bibinfo{year}{2007}.
\newblock \bibinfo{title}{Coupling level set/vof/ghost fluid methods:
  Validation and application to 3d simulation of the primary break-up of a
  liquid jet}.
\newblock \bibinfo{journal}{International Journal of Multiphase Flow}
  \bibinfo{volume}{33}, \bibinfo{pages}{510 -- 524}.
\newblock
  \DOIprefix\doi{https://doi.org/10.1016/j.ijmultiphaseflow.2006.11.001}.
\bibitem[{Navarro-Martinez(2014)}]{NavarroMartinez2014}
\bibinfo{author}{Navarro-Martinez, S.}, \bibinfo{year}{2014}.
\newblock \bibinfo{title}{Large eddy simulation of spray atomization with a
  probability density function method}.
\newblock \bibinfo{journal}{International Journal of Multiphase Flow}
  \bibinfo{volume}{63}, \bibinfo{pages}{11--22}.
\bibitem[{Olsson and Kreiss(2005)}]{Olsson2005}
\bibinfo{author}{Olsson, E.}, \bibinfo{author}{Kreiss, G.},
  \bibinfo{year}{2005}.
\newblock \bibinfo{title}{A conservative level set method for two phase flow}.
\newblock \bibinfo{journal}{Journal of Computational Physics}
  \bibinfo{volume}{210}, \bibinfo{pages}{225--246}.
\newblock \DOIprefix\doi{https://doi.org/10.1016/j.jcp.2005.04.007}.
\bibitem[{Olsson et~al.(2007)Olsson, Kreiss and Zahedi}]{Olsson2007}
\bibinfo{author}{Olsson, E.}, \bibinfo{author}{Kreiss, G.},
  \bibinfo{author}{Zahedi, S.}, \bibinfo{year}{2007}.
\newblock \bibinfo{title}{A conservative level set method for two phase flow
  ii}.
\newblock \bibinfo{journal}{Journal of Computational Physics}
  \bibinfo{volume}{225}, \bibinfo{pages}{785--807}.
\bibitem[{Osher and Fedkiw(2006)}]{Osher2006}
\bibinfo{author}{Osher, S.}, \bibinfo{author}{Fedkiw, R.},
  \bibinfo{year}{2006}.
\newblock \bibinfo{title}{Level set methods and dynamic implicit surfaces}.
  volume \bibinfo{volume}{153}.
\newblock \bibinfo{publisher}{Springer Science \& Business Media}.
\bibitem[{Osher and Sethian(1988)}]{Osher1988}
\bibinfo{author}{Osher, S.}, \bibinfo{author}{Sethian, J.A.},
  \bibinfo{year}{1988}.
\newblock \bibinfo{title}{Fronts propagating with curvature-dependent speed:
  Algorithms based on hamilton-jacobi formulations}.
\newblock \bibinfo{journal}{Journal of Computational Physics}
  \bibinfo{volume}{79}, \bibinfo{pages}{12--49}.
\newblock \DOIprefix\doi{https://doi.org/10.1016/0021-9991(88)90002-2}.
\bibitem[{Pope(2001)}]{pope2001turbulent}
\bibinfo{author}{Pope, S.B.}, \bibinfo{year}{2001}.
\newblock \bibinfo{title}{Turbulent flows}.
\bibitem[{Pozzetti and Peters(2018)}]{Pozzetti2018}
\bibinfo{author}{Pozzetti, G.}, \bibinfo{author}{Peters, B.},
  \bibinfo{year}{2018}.
\newblock \bibinfo{title}{A multiscale dem-vof method for the simulation of
  three-phase flows}.
\newblock \bibinfo{journal}{International Journal of Multiphase Flow}
  \bibinfo{volume}{99}, \bibinfo{pages}{186 -- 204}.
\newblock
  \DOIprefix\doi{https://doi.org/10.1016/j.ijmultiphaseflow.2017.10.008}.
\bibitem[{Prosperetti and Tryggvason(2009)}]{Prosperetti2009}
\bibinfo{author}{Prosperetti, A.}, \bibinfo{author}{Tryggvason, G.},
  \bibinfo{year}{2009}.
\newblock \bibinfo{title}{Computational Methods for Multiphase Flow}.
\newblock \bibinfo{publisher}{Cambridge University Press}.
\bibitem[{Qian et~al.(2018)Qian, Wei and Xiao}]{Qian2018}
\bibinfo{author}{Qian, L.}, \bibinfo{author}{Wei, Y.}, \bibinfo{author}{Xiao,
  F.}, \bibinfo{year}{2018}.
\newblock \bibinfo{title}{Coupled thinc and level set method: A conservative
  interface capturing scheme with high-order surface representations}.
\newblock \bibinfo{journal}{Journal of Computational Physics}
  \bibinfo{volume}{373}, \bibinfo{pages}{284--303}.
\bibitem[{Rider and Kothe(1998)}]{Rider1998}
\bibinfo{author}{Rider, W.J.}, \bibinfo{author}{Kothe, D.B.},
  \bibinfo{year}{1998}.
\newblock \bibinfo{title}{Reconstructing volume tracking}.
\newblock \bibinfo{journal}{Journal of computational physics}
  \bibinfo{volume}{141}, \bibinfo{pages}{112--152}.
\bibitem[{Roenby et~al.(2016)Roenby, Bredmose and Jasak}]{Roenby2016}
\bibinfo{author}{Roenby, J.}, \bibinfo{author}{Bredmose, H.},
  \bibinfo{author}{Jasak, H.}, \bibinfo{year}{2016}.
\newblock \bibinfo{title}{A computational method for sharp interface
  advection}.
\newblock \bibinfo{journal}{Royal Society open science} .
\bibitem[{Sandberg et~al.(2019)Sandberg, Hattel and Spangenberg}]{Sandberg2019}
\bibinfo{author}{Sandberg, M.}, \bibinfo{author}{Hattel, J.H.},
  \bibinfo{author}{Spangenberg, J.}, \bibinfo{year}{2019}.
\newblock \bibinfo{title}{Simulation of liquid composite moulding using a
  finite volume scheme and the level-set method}.
\newblock \bibinfo{journal}{International Journal of Multiphase Flow}
  \bibinfo{volume}{118}, \bibinfo{pages}{183 -- 192}.
\newblock
  \DOIprefix\doi{https://doi.org/10.1016/j.ijmultiphaseflow.2019.06.014}.
\bibitem[{Scardovelli and Zaleski(1999)}]{Scardovelli1999}
\bibinfo{author}{Scardovelli, R.}, \bibinfo{author}{Zaleski, S.},
  \bibinfo{year}{1999}.
\newblock \bibinfo{title}{Direct numerical simulation of free-surface and
  interfacial flow}.
\newblock \bibinfo{journal}{Annual review of fluid mechanics}
  \bibinfo{volume}{31}, \bibinfo{pages}{567--603}.
\bibitem[{Scardovelli and Zaleski(2000)}]{Scardovelli2000}
\bibinfo{author}{Scardovelli, R.}, \bibinfo{author}{Zaleski, S.},
  \bibinfo{year}{2000}.
\newblock \bibinfo{title}{Analytical relations connecting linear interfaces and
  volume fractions in rectangular grids}.
\newblock \bibinfo{journal}{Journal of Computational Physics}
  \bibinfo{volume}{164}, \bibinfo{pages}{228--237}.
\bibitem[{Sethian(1996)}]{Sethian1996}
\bibinfo{author}{Sethian, J.A.}, \bibinfo{year}{1996}.
\newblock \bibinfo{title}{Level set methods, evolving interfaces in geometry,
  fluid mechanics comuputer vision, and materials sciences}.
\newblock \bibinfo{journal}{Cambridge Monographs on Applied and Computational
  Mathematics, 3} .
\bibitem[{Sethian(1999)}]{Sethian1999}
\bibinfo{author}{Sethian, J.A.}, \bibinfo{year}{1999}.
\newblock \bibinfo{title}{Level set methods and fast marching methods: evolving
  interfaces in computational geometry, fluid mechanics, computer vision, and
  materials science}. volume~\bibinfo{volume}{3}.
\newblock \bibinfo{publisher}{Cambridge university press}.
\bibitem[{Sussman and Puckett(2000)}]{Sussman2000}
\bibinfo{author}{Sussman, M.}, \bibinfo{author}{Puckett, E.G.},
  \bibinfo{year}{2000}.
\newblock \bibinfo{title}{A coupled level set and volume-of-fluid method for
  computing 3d and axisymmetric incompressible two-phase flows}.
\newblock \bibinfo{journal}{Journal of Computational Physics}
  \bibinfo{volume}{162}, \bibinfo{pages}{301--337}.
\newblock \DOIprefix\doi{https://doi.org/10.1006/jcph.2000.6537}.
\bibitem[{Tanguy and Berlemont(2005)}]{Tanguy2005}
\bibinfo{author}{Tanguy, S.}, \bibinfo{author}{Berlemont, A.},
  \bibinfo{year}{2005}.
\newblock \bibinfo{title}{Application of a level set method for simulation of
  droplet collisions}.
\newblock \bibinfo{journal}{International Journal of Multiphase Flow}
  \bibinfo{volume}{31}, \bibinfo{pages}{1015 -- 1035}.
\newblock
  \DOIprefix\doi{https://doi.org/10.1016/j.ijmultiphaseflow.2005.05.010}.
\bibitem[{Toro(1997)}]{Toro1997}
\bibinfo{author}{Toro, E.F.}, \bibinfo{year}{1997}.
\newblock \bibinfo{title}{Splitting schemes for pdes with source terms}, in:
  \bibinfo{booktitle}{Riemann Solvers and Numerical Methods for Fluid
  Dynamics}. \bibinfo{publisher}{Springer}, pp. \bibinfo{pages}{497--507}.
\bibitem[{Toro and Titarev(2005)}]{toro2005tvd}
\bibinfo{author}{Toro, E.F.}, \bibinfo{author}{Titarev, V.A.},
  \bibinfo{year}{2005}.
\newblock \bibinfo{title}{Tvd fluxes for the high-order ader schemes}.
\newblock \bibinfo{journal}{Journal of Scientific Computing}
  \bibinfo{volume}{24}, \bibinfo{pages}{285--309}.
\bibitem[{Tryggvason et~al.(2011)Tryggvason, Scardovelli and
  Zaleski}]{Tryggvason2011}
\bibinfo{author}{Tryggvason, G.}, \bibinfo{author}{Scardovelli, R.},
  \bibinfo{author}{Zaleski, S.}, \bibinfo{year}{2011}.
\newblock \bibinfo{title}{Direct numerical simulations of gas--liquid
  multiphase flows}.
\newblock \bibinfo{publisher}{Cambridge University Press}.
\bibitem[{Vallet and Borghi(1999)}]{Vallet1999}
\bibinfo{author}{Vallet, A.}, \bibinfo{author}{Borghi, R.},
  \bibinfo{year}{1999}.
\newblock \bibinfo{title}{Modelisation eulerienne de i’atomisation d’un jet
  liquide}.
\newblock \bibinfo{journal}{C. R. Acad. Sci. Paris, t. 327} \bibinfo{volume}{t.
  327}, \bibinfo{pages}{1115--1200}.
\bibitem[{Vallet et~al.(2001)Vallet, Burluka and Borghi}]{Vallet2001}
\bibinfo{author}{Vallet, A.}, \bibinfo{author}{Burluka, A.},
  \bibinfo{author}{Borghi, R.}, \bibinfo{year}{2001}.
\newblock \bibinfo{title}{Development of an eulerian model for the
  “atomization” of a liquid jet}.
\newblock \bibinfo{journal}{Atomization and Sprays} \bibinfo{volume}{11},
  \bibinfo{pages}{619--642}.
\bibitem[{Wang et~al.(2009)Wang, Yang, Koo and Stern}]{Wang2009}
\bibinfo{author}{Wang, Z.}, \bibinfo{author}{Yang, J.}, \bibinfo{author}{Koo,
  B.}, \bibinfo{author}{Stern, F.}, \bibinfo{year}{2009}.
\newblock \bibinfo{title}{A coupled level set and volume-of-fluid method for
  sharp interface simulation of plunging breaking waves}.
\newblock \bibinfo{journal}{International Journal of Multiphase Flow}
  \bibinfo{volume}{35}, \bibinfo{pages}{227 -- 246}.
\newblock
  \DOIprefix\doi{https://doi.org/10.1016/j.ijmultiphaseflow.2008.11.004}.
\bibitem[{Weller et~al.(1998)Weller, Tabor, Jasak and Fureby}]{Weller1998}
\bibinfo{author}{Weller, H.G.}, \bibinfo{author}{Tabor, G.},
  \bibinfo{author}{Jasak, H.}, \bibinfo{author}{Fureby, C.},
  \bibinfo{year}{1998}.
\newblock \bibinfo{title}{A tensorial approach to computational continuum
  mechanics using object-oriented techniques}.
\newblock \bibinfo{journal}{Comput. Phys.} \bibinfo{volume}{12},
  \bibinfo{pages}{620--631}.
\newblock \DOIprefix\doi{10.1063/1.168744}.
\bibitem[{Xiao et~al.(2011)Xiao, Ii and Chen}]{Xiao2011}
\bibinfo{author}{Xiao, F.}, \bibinfo{author}{Ii, S.}, \bibinfo{author}{Chen,
  C.}, \bibinfo{year}{2011}.
\newblock \bibinfo{title}{Revisit to the thinc scheme: a simple algebraic vof
  algorithm}.
\newblock \bibinfo{journal}{Journal of Computational Physics}
  \bibinfo{volume}{230}, \bibinfo{pages}{7086--7092}.
\bibitem[{Xie and Xiao(2017)}]{Xie2017}
\bibinfo{author}{Xie, B.}, \bibinfo{author}{Xiao, F.}, \bibinfo{year}{2017}.
\newblock \bibinfo{title}{Toward efficient and accurate interface capturing on
  arbitrary hybrid unstructured grids: The thinc method with quadratic surface
  representation and gaussian quadrature}.
\newblock \bibinfo{journal}{Journal of Computational Physics}
  \bibinfo{volume}{349}, \bibinfo{pages}{415--440}.
\bibitem[{Yokoi(2007)}]{Yokoi2007}
\bibinfo{author}{Yokoi, K.}, \bibinfo{year}{2007}.
\newblock \bibinfo{title}{Efficient implementation of thinc scheme: a simple
  and practical smoothed vof algorithm}.
\newblock \bibinfo{journal}{Journal of Computational Physics}
  \bibinfo{volume}{226}, \bibinfo{pages}{1985--2002}.
\bibitem[{Zhao et~al.(2014)Zhao, Mao, Liu, Bai and Willims}]{Zhao2014}
\bibinfo{author}{Zhao, L.H.}, \bibinfo{author}{Mao, J.}, \bibinfo{author}{Liu,
  X.Q.}, \bibinfo{author}{Bai, X.}, \bibinfo{author}{Willims, J.},
  \bibinfo{year}{2014}.
\newblock \bibinfo{title}{Improved conservative level set method for free
  surface flow simulation}.
\newblock \bibinfo{journal}{Journal of Hydrodynamics, Ser. B}
  \bibinfo{volume}{26}, \bibinfo{pages}{316 -- 325}.
\newblock \DOIprefix\doi{https://doi.org/10.1016/S1001-6058(14)60035-4}.
\bibitem[{Zhao and Chen(2017)}]{Zhao2017}
\bibinfo{author}{Zhao, Y.}, \bibinfo{author}{Chen, H.C.}, \bibinfo{year}{2017}.
\newblock \bibinfo{title}{A new coupled level set and volume-of-fluid method to
  capture free surface on an overset grid system}.
\newblock \bibinfo{journal}{International Journal of Multiphase Flow}
  \bibinfo{volume}{90}, \bibinfo{pages}{144 -- 155}.
\newblock
  \DOIprefix\doi{https://doi.org/10.1016/j.ijmultiphaseflow.2017.01.002}.

\end{thebibliography}

\end{document}